\documentclass[a4paper,12pt]{article}
\usepackage{jheppub} 
\pdfoutput=1 
\usepackage{hyperref}
\usepackage{graphicx}

         \let\g = \gamma     \let\e = \epsilon

\newcommand{\nn}{\nonumber}

\newcommand{\beq}{\begin{equation}}
\newcommand{\eeq}{\end{equation}}
\newcommand{\bea}{\begin{eqnarray}}
\newcommand{\eea}{\end{eqnarray}}
\newcommand{\beqa}{\begin{eqnarray}}
\newcommand{\eeqa}{\end{eqnarray}}

\newcommand{\AL}[1]{\langle#1|}
\newcommand{\AR}[1]{|#1\rangle}
\newcommand{\SL}[1]{[#1|}
\newcommand{\SR}[1]{|#1]}
\renewcommand{\AA}[1]{\langle#1\rangle}
\newcommand{\SSS}[1]{[#1]}
\newcommand{\AS}[1]{\langle#1]}

\newcommand{\lid}[2]{#1\!\cdot\!#2}
\newcommand{\slashk}{k \! \! \!  /}
\newcommand{\slashK}{K \! \! \!  /}
\newcommand{\slashp}{p \! \! \!  /}

\newcommand{\slashr}{r \! \! \!  /}
\newcommand{\slashe}{e \! \! \!  /}
\newcommand{\slasheps}{\epsilon \! \! \!  /}
\newcommand{\kapp}{\kappa}
\newcommand{\kapphat}{\hat{\kappa}}
\newcommand{\kstr}{\kappa^*}
\newcommand{\kstrhat}{\hat{\kappa}^*}
\newcommand{\qb}{\bar{q}}
\newcommand{\ub}{\bar{u}}
\newcommand{\db}{\bar{d}}
\newcommand{\Amp}{\mathcal{A}}

\newcommand{\imag}{\mathrm{i}}

\newcommand{\graph}[3]{\raisebox{-#3ex}{\epsfig{file=figures/#1.pdf,width=#2ex}}}

\title{\boldmath QCD amplitudes with 2 initial spacelike legs via generalised BCFW recursion}

\author{Krzysztof Kutak,}
\author{Andreas van Hameren}
\author{and Mirko Serino}
\affiliation{The H. Niewodnicza\'nski Institute of Nuclear Physics, Polish Academy of Sciences, \\
ul. Radzikowskiego 152, 31-342, Cracow, Poland}

\emailAdd{krzysztof.kutak@ifj.edu.pl}
\emailAdd{andre.hameren@ifj.edu.pl}
\emailAdd{mirko.serino@ifj.edu.pl.edu}

\abstract{
We complete the generalisation of the BCFW recursion relation to the off-shell case, 
allowing for the computation of tree level scattering amplitudes for full High Energy Factorisation (HEF), 
i.e.\ with both incoming partons having a non-vanishing transverse momentum. 
We provide explicit results for color-ordered amplitudes with two off-shell legs in massless QCD
up to 4 point, continuing the program begun in two previous papers.
For the 4-fermion amplitudes, which are not BCFW-recursible,
we perform a diagrammatic computation, so as to offer a complete set of expressions.
We explicitly show and discuss some plots of the squared $2 \rightarrow 2$ matrix elements
as functions of the differences in rapidity and azimuthal angle of the final state particles.
}

\begin{document}

\begin{flushright}
 IFJPAN-IV-2016-27 \\
\end{flushright}

\maketitle
\flushbottom

\section{Introduction}

It is really remarkable how simple QCD tree level scattering amplitudes are, if compared to the impressive
number of Feynman diagrams required to evaluate even the simplest of them.
This complication lies, on one hand, 
in the non-abelian nature of the theory, i.e. in the complicated color structure of QCD; 
secondly, in the fact that the ordinary Feynman diagrams employ unphysical, i.e. gauge-dependent degrees of freedom.

Great progress came already with the introduction of approaches to group Feynman diagrams 
into gauge-invariant subsets~\cite{Cvitanovic:1980bu,DelDuca:1999rs}, which naturally goes hand in hand with employing the spinor helicity formalism to get rid of gauge 
redundancy~\cite{Kleiss:1985yh,Gunion:1985vca,Xu:1986xb}.
In combination with off-shell recursions~\cite{Berends:1987me,Mangano:1987xk,Mangano:1990by,Berends:1989hf,Kosower:1989xy},
these novel techniques allowed to reach compact expressions for many scattering amplitudes.

The discovery of a purely on-shell recursion by Britto, Cachazo, Feng and Witten (BCFW)~\cite{ Britto:2004ap,Britto:2005fq}
showed that the occurrence of very compact expressions for multi-gluon amplitudes is indeed rather natural,
since it is a recursion of gauge invariant compact expressions themselves, rather than of gauge non-invariant vertices or off-shell currents.
This recursion, originally introduced only for purely gluon amplitudes, was soon generalised to many other theories, 
in particular to amplitudes involving massless quark-antiquark pairs~\cite{Luo:2005rx}.

Until very recently, all this progress had benefitted the computation of only purely on-shell amplitudes, which find their natural application
in the collinear factorization framework. Nevertheless, there are energy regimes where other factorization schemes prove to be
more appropriate. For the purpose of this paper, we focus on the prescriptions demanding off-shell partonic initial states, 
like for example High Energy Factorization (HEF)~\cite{Catani:1990eg,Collins:1991ty,Deak:2009xt} and the parton Reggeization approach~\cite{Fadin:1993wh,Fadin:1996nw}.
Recent applications of such amplitudes to high-energy phenomenology can be found in~\cite{Nefedov:2013ywa,Kniehl:2014qva,Nefedov:2014qea,
Karpishkov:2014epa,Kutak:2016mik,Kutak:2016ukc}.

On one hand, amplitudes with off-shell partons are known for the long-standing problem of gauge invariance, 
i.e.\ the existence of Ward identities for the on-shell gluons and the possibility to choose an arbitrary gauge for the internal gluon propagators is non trivial. 
Nevertheless, this has been completely solved at tree level, most recently in~\cite{vanHameren:2012if,vanHameren:2013csa}, 
which employ techniques based on the spinor helicity formalism, after the pioneering work of Lipatov and his collaborators~\cite{Lipatov:1995pn,Lipatov:2000se,Antonov:2004hh}.
Some more formal insight into the problem is coming, on the other hand, from Wilson-lines based approaches~\cite{Kotko:2014aba,Kotko:2016qxv} and, 
very recently, from attempts to generalise the Grassmanian construction to $\mathcal{N}=4$ Super Yang-Mills~\cite{Bork:2016xfn,Bork:2016egt}.

On the other hand, the bottleneck for these off-shell amplitudes is even more severe than in the on-shell case, because, when off-shell partons
are present,  gauge invariance comes at the price of an increased number of Feynman diagrams. 
However, generalisations of the BCFW recursion have been proved for multi-gluon amplitudes with off-shell gluons and amplitudes with 
one off-shell particle and 1 fermion pair on the external lines~\cite{vanHameren:2014iua,vanHameren:2015bba}.
In this paper, we set out to close the circle and prove a generalisation of the BCFW recursion relation which holds for any amplitude
with up to two off-shell particles which features at least one gluon on the external lines.
This generalises the results of~\cite{ArkaniHamed:2008yf,Cheung:2008dn} to the High Energy Factorization case.

This paper is organized as follows.
In section \ref{2_off_legs} we formally prove that it is possible to apply the BCFW recursion to tree level QCD 
scattering amplitudes with at least one gluon with up to two off-shell legs,  using straightforward diagrammatic arguments.
In section \ref{3_Point_Amps} we compute all the 3-point  amplitudes with two off-shell legs, which are the fundamental starting point of the
recursion and can be obtained via BCFW itself starting from the known on-shell results, as well as diagrammatically.
Section \ref{4_Point_Amps} presents the explicit derivations of the 4 point amplitudes for $2\rightarrow 2$ 
scattering in High Energy Factorization. 
Their BCFW calculation is worked out. For the cases with 4 fermions, when BCFW does not work, 
a diagrammatic evaluation is performed, so as to explicitly compute them all. Section \ref{MHV} collects the all-leg results for MHV amplitudes.
Finally, section \ref{Squared} shows an explicit analytical study of 
the squared matrix elements and how these compare to the Hybrid Factorization case for the 4-gluon scattering.   
Some more technical details and cross checks are presented in the appendices.

\section{The BCFW relation for amplitudes with two off-shell legs}\label{2_off_legs}

We always consider scattering amplitudes with all particles outgoing or incoming
and refer the reader to~\cite{vanHameren:2014iua,vanHameren:2015bba} for the details
about our conventions as well as an introduction to the BCFW recursion.
General informations about the notations are contained in Appendix~\ref{AppSpinors}. 

\subsection{Recursibility with two off-shell legs}

In order to apply the BCFW recursion, one has to make the scattering amplitude $\Amp$  
a function of the auxiliary complex variable $\Amp(z)$ in such a way that
\begin{equation}
\lim_{z\to\infty} \Amp(z) = 0 \, .
\end{equation}
We refer to ~\cite{Britto:2005fq,Luo:2005rx} for the original proofs in the on-shell case respectively for only gluons and with fermions as well. 
In~\cite{vanHameren:2015bba} a thorough discussion of allowed shifts for amplitudes with only one off-shell leg can be found.

In this section we show that amplitudes with two off-shell legs and any number of fermion pairs are always BCFW-recursible,
as long as there is at least one gluon among the external particles.
By this we mean that they can always be expressed in terms of lower point amplitudes and of amplitudes of the same rank but with one less off-shell leg, 
according to the classification of residues provided in \cite{vanHameren:2014iua,vanHameren:2015bba}.
An argument leading to the corresponding conclusions for on-shell amplitudes has been around for some time \cite{Cheung:2008dn}.
It employs the background field method, previously used in \cite{ArkaniHamed:2008yf} to give a physical motivation of the existence of
the BCFW recursion relation. Its conclusion is that BCFW is viable for gauge theory and gravity amplitudes coupled to fermions and/or scalars
as long as at least 1 gluon or graviton is an external leg \cite{Cheung:2008dn}.
Here it will be sufficient for us to argue in terms of individual diagrams.

In the gluon case, one can always shift the two off-shell legs and thus we get a factor $~1/z^2$ from the external lines, as in the case which 
is diagrammatically examined in \cite{Britto:2005fq}, when the shifted gluons have helicities $ (h_i,h_j) = (-,+)$.
In fact, as off-shell gluons do not have corresponding polarisation vectors, we deliberately include
their propagators in the amplitude in order to ensure proper behaviour for $z\rightarrow\infty$. 

This implies the existence of additional poles, dubbed `C' and `D' terms, whose residues are just the amplitude 
with the leg exhibiting the pole being properly taken on-shell,
as detailed in \cite{vanHameren:2014iua,vanHameren:2015bba}.
Another kind of poles featured only by off-shell amplitudes are due to vanishing denominators in eikonal propagators.

\begin{figure}[ht]
\begin{center}
\includegraphics[scale=0.6]{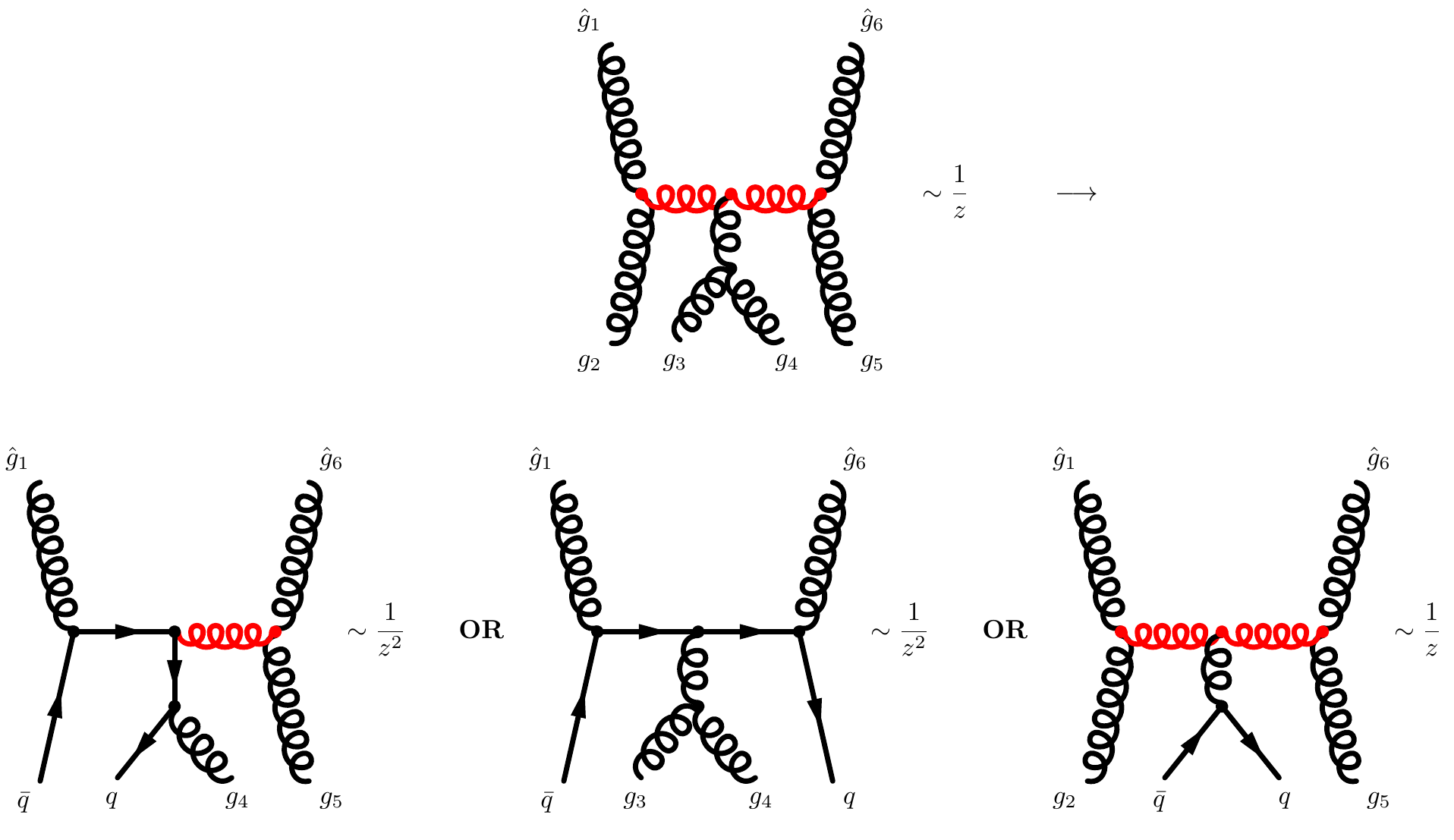}
\caption{Example of the procedure to infer, from the asymptotic of diagrams for an $n$-gluon amplitude,
the behaviour of diagrams for an amplitude with $(n-2)$ gluon and 1 fermion pair.
Highlighted in red are the vertices and propagators through which $z$ flows.}
\label{shift_1_fig}
\end{center}
\end{figure}

From the original argument in \cite{Britto:2005fq} for all-gluon amplitudes, we know that in individual diagrams 
there can be at most one more three-gluon vertex than there are propagators. 
This implies that, as long as we have a factor $1/z^2$ from external legs, 
all diagrams behave at least as well as $\sim 1/z$ asymptotically.
Also, from the observations above we understand that, as long as there is one gluon among the external particles, 
we can get a factor $~1/z$, whether it is on-shell or off-shell. 
The question is whether one can explicitly prove that this is enough in all cases.
Fermions interact with the gauge field via a 3-point momentum-independent 
vertex and their propagators approach a constant for $z \rightarrow \infty$.
The key observation is that $n$-point amplitudes with $k$ fermion pairs and $m$ gluons ($ n = 2\,k +m $)
can be thought of as $n$-point gluon amplitudes made by diagrams obtained by turning $k$ couples of external gluons into fermions 
and changing accordingly vertices and propagators along the diagram, as depicted in the example in Fig. \ref{shift_1_fig}.

\begin{figure}[ht]
\begin{center}
\includegraphics[scale=0.6]{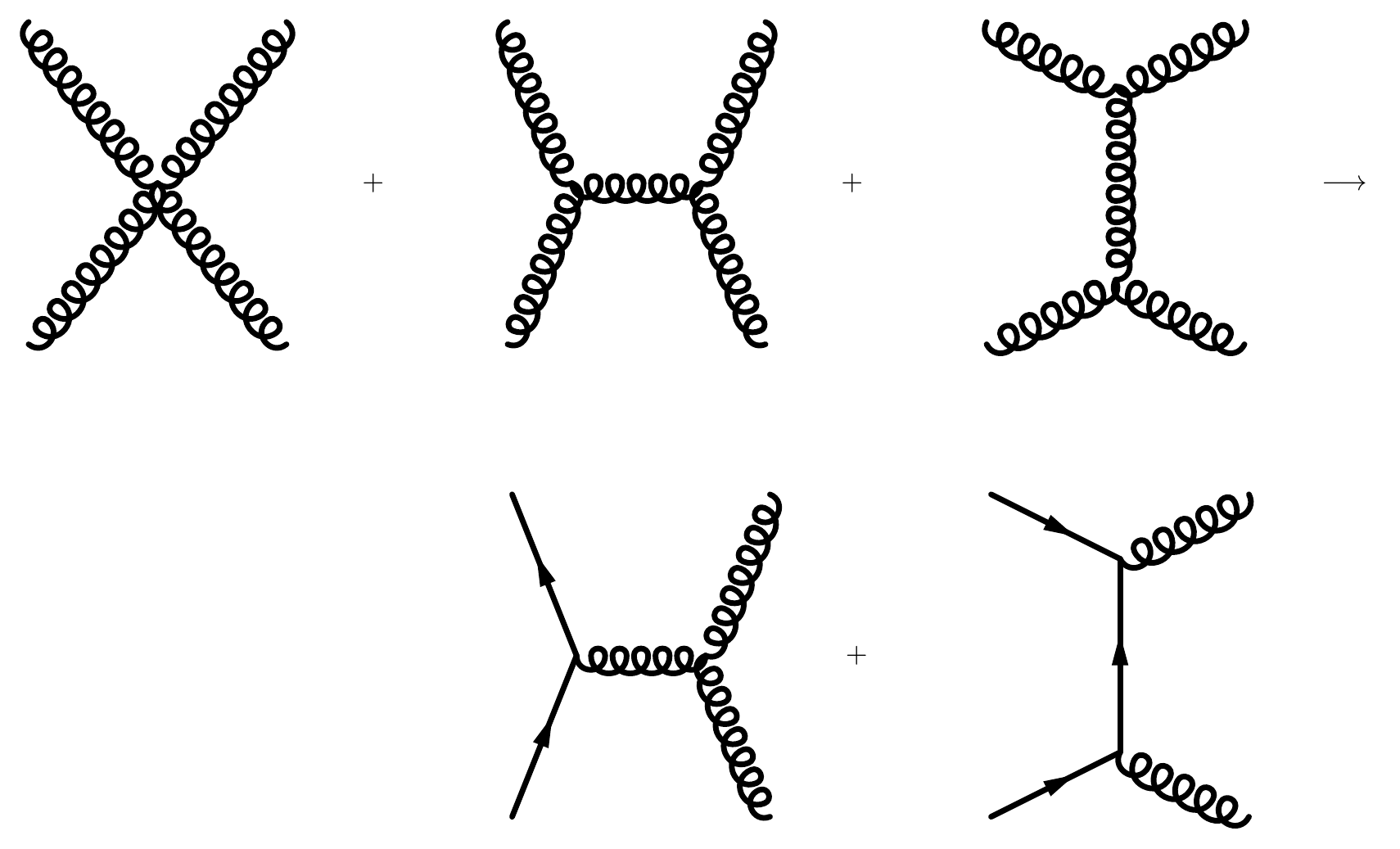}
\caption{How to get the planar contributions to the scattering of $2$-gluon and 2 fermions
from the $2 \rightarrow 2$ gluon scattering}
\label{from_g_to_f}
\end{center}
\end{figure}

Clearly, we can do this switch only for the diagrams for which it is possible to make the fermion number 
flow along the diagram without meeting a 4-point vertex, 
but this simply means that the topologies of diagrams for $n$-gluon scattering for which this is not possible
do not contribute to the amplitudes describing scattering of $n-2k$ gluons and $2k$ fermions.
For instance, in the case of $4$-point amplitudes, 
we can obtain the color ordered amplitudes for $g g \psi \bar\psi$ from those for $4$-gluon scattering 
with this prescription and, clearly, the diagram made by the 4-gluon vertex will not generate any contribution.
This is illustrated in Fig. \ref{from_g_to_f}, where the three planar diagrams contributing to the 4-gluon 
amplitude are turned into the two diagrams representing a scattering process involving two fermions and two gluons.

Now, for every possible Feynman diagram, the numbers of fermions and gluons, $N_{f/g}$, are related
to the number of propagators and vertices by the relations
\bea
N_f &=& 2\, \left( V_f - P_f \right) \, , 
\label{f_num} \\
N_g &=& V_f + 3\, V_{3g} + 4\, V_{4g} - 2\, P_g  \, ,
\label{g_nums}
\eea
where the meaning of subscripts is obvious.

Let us suppose that an $n$-gluon amplitude is turned into an amplitude with $n-2k$ gluons and
$k$ fermion pairs switching vertices and propagators according to $P_g \rightarrow P_f$
and $V_{3g} \rightarrow V_f$ along the flow of the fermion number.
Then we can express $n$ in two ways for diagrams of the two amplitudes,
\begin{equation}
n = 
\left\{ \begin{array}{l}
3\,V_{3g} + 4\,V_{4g} - 2\, P_g
\\
3\,V'_{f} + 3\,V'_{3g} + 4\,V'_{4g} - 2\, \left( P'_f + P'_g \right)
\\
\end{array} \right.
\label{balance}
\end{equation}
where primes of course refer to the amplitude with fermions. \\
As, obviously, $V'_{4g} = V_{4g}$ and $V_{3g} = V'_{3g} + V'_f $, we infer that
\beq
2\,\Delta V_{3g} - 2\,\Delta P_g = 2\, \left( V'_f - P'_f \right)  = 2k \Rightarrow \Delta \left( V_{3g} - P_{g} \right)  = k > 0  \, ,
\eeq
where we have used (\ref{f_num}) and $\Delta$ is the difference between non-primed and primed quantities.
The last equation means that the difference between gluon vertices and gluon propagators is bigger for the 
all-gluon amplitude than for the amplitude with $2k$ fermions. 
Consequently, the asymptotic behaviour of internal lines of the latter as a function of $z$ will always be better, i.e. at least as good as $~ O(z^0)$
As we can always shift one gluon leg which goes as $~ 1/z$ for $z \to \infty$, the desired asymptotic condition is fulfilled. \\
We conclude by pointing out that eikonal propagators behave as gluon propagators for $z\rightarrow \infty$,
as the direction $p$ does not shift for off-shell particles, whereas eikonal vertices do not depend on the momenta of the eikonal particles 
(see the Feynman rules in~\cite{vanHameren:2015bba}),
so that our arguments hold for diagrams containing such terms too.

Finally, in the event that one has to shift one gluon and one fermion line,
it is always possible to do that without getting a $~ z$ factor from the fermion spinor.
For instance, it is known in the on-shell case that
it is not allowed to shift one fermion and one gluon with the same helicity sign. 
This happens because an amplitude with, for instance, one fermion pair is given by
\begin{equation}
\Amp(\qb^+,q^-,g_1,\dots, g_n) =  \AL{q} \dots \SR{\qb} \,  ,
\label{Proper_spinors_1}
\end{equation}
If one shifts the fermion $q^-$ and one gluon with negative helicity $g^-$, 
then the shift vector guaranteeing a $1/z$ factor from the gluon is 
$e^\mu = \AL{g} \g^\mu \SR{q}/2 $, for which the large $z$ behaviours of the legs cancel out.
On the other hand, $\AL{g} \g^\mu \SR{\qb}/2$ is manifestly a good shift vector.
A completely similar argument holds for $\Amp(\qb^-,q^+,g_1,\dots, g_n) =  \SL{q} \dots \AR{\qb}$.
Then, when working with amplitudes with on-shell fermions, one can always choose the shift vector
in such a way that the external fermion spinor does not shift. 
In \cite{vanHameren:2015bba} we called this the {\em original LW prescription} because it was first discussed in \cite{Luo:2005rx}.
Finally, if the fermion is off-shell, then the spinor is not shifted anyway, so that our argument is complete.
This concludes our proof.

\subsection{Residues}\label{res}

The classification of the possible residues was already presented in \cite{vanHameren:2014iua,vanHameren:2015bba}.
The general formula which generalizes the BCFW recursion relation to the off-shell case is
\beq
\Amp(0) = \sum_{s=g,f} \left(  \sum_{p} \sum_{h=+,-} \mathrm{A}^s_{p,h} + \sum_{i} \mathrm{B}^s_i  + \mathrm{C}^s + \mathrm{D}^s \right) \, ,
\eeq
where the index $s$ refers to the particle species, which can be gluon $(g)$ or fermion $(f)$, and $h$ refers to the helicity.
We recall the nature of these contributions.
The $\mathrm{A}^{s}_{p,h}$ residues are due to the poles which appear also in the original BCFW recursion for on-shell amplitudes, whereas
the $\mathrm{B}^{s}_i$ terms are due to the poles appearing in the propagator of auxiliary eikonal quarks~\cite{vanHameren:2012if,vanHameren:2013csa}.
If the $i$-th particle is on-shell or off-shell but shifted, these terms are not present.
The case when one off-shell leg is right next to a shifted leg leads to the appearance
of $B$ residues, which imply 2-point functions defined only for shifted momenta. 
Their proper treatment boils down to the  the same final prescription, no matter whether the off-shell particle is a gluon or a fermion.
In the case it is a gluon, this problem was already addressed in \cite{vanHameren:2014iua}.
Here we collect the results in order to have them all available in the same place.
Notice that the $k$-lines can represent an off-shell gluon as well as an off-shell fermion and the associated eikonal propagators.
We explicitly distinguish the cases in which the $i$-th particle in the 2-point function is on-shell, as well as the shift vector;
instead, the double $j$-line can be either on-shell or off-shell, but the resulting different shifts are specified at the bottom.
\bea
&&
\graph{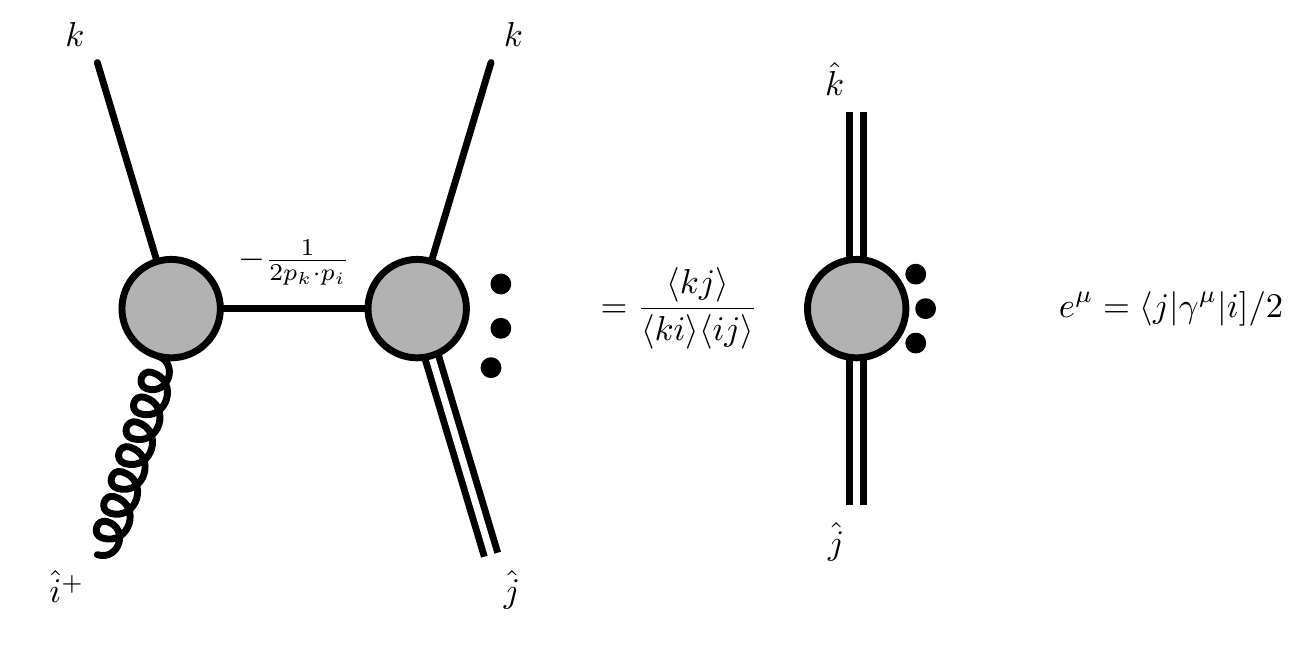}{60}{10}
\nn \\
&&
\hat{\kapp}_k = \frac{\AL{j} \slashk_k + \slashp_i \SR{k}}{\AA{jk}} \, , 
\quad
\hat{\kapp}_j = \frac{\AL{k} \slashk_j + \slashp_i \SR{j}}{\AA{kj}} \, \lor \,  \SR{\hat{j}} = \frac{ \left( \slashp_i+\slashp_j \right)\AR{k}}{\AA{jk}}
 \nn
\eea

\bea
&&
\graph{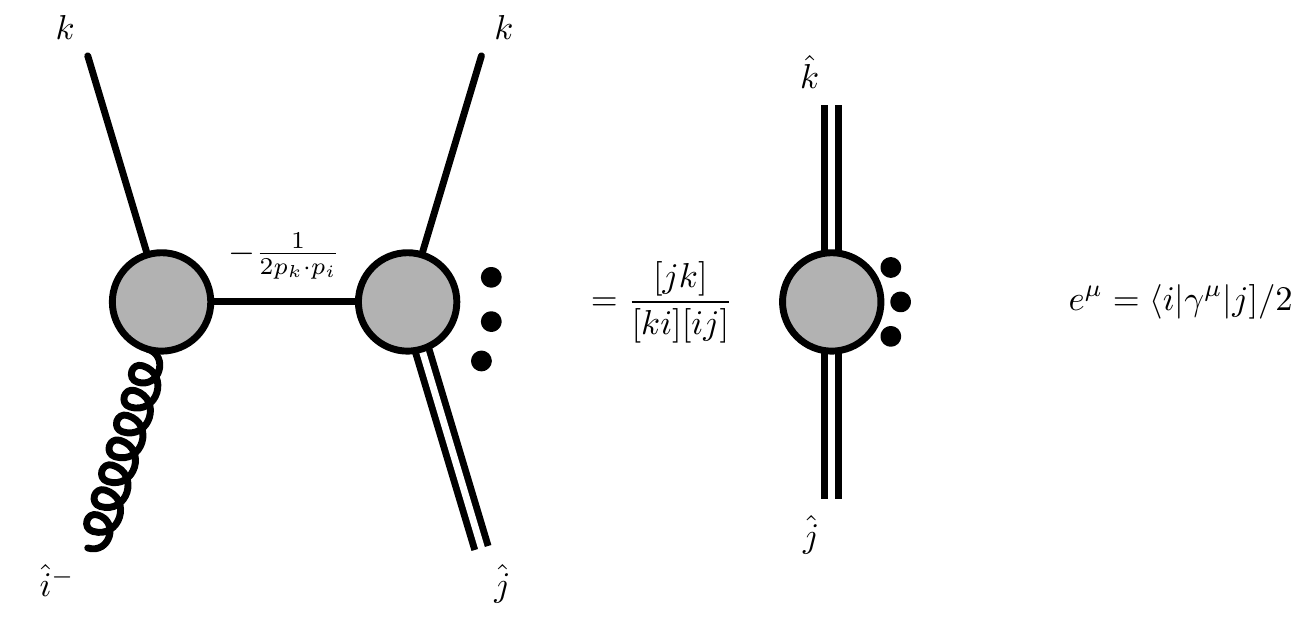}{60}{10}
\nn \\
&&
\hat{\kstr}_k = \frac{\AL{k} \slashk_k + \slashp_i \SR{j}}{\SSS{kj}} \, , 
\quad
\hat{\kstr}_j = \frac{\AL{j} \slashk_j + \slashp_i \SR{k}}{\SSS{jk}} \, \lor \,  \SR{\hat{j}} = \frac{ \left( \slashp_i+\slashp_j \right)\SR{k}}{\SSS{jk}}
\nn
\eea

\bea
&&
\graph{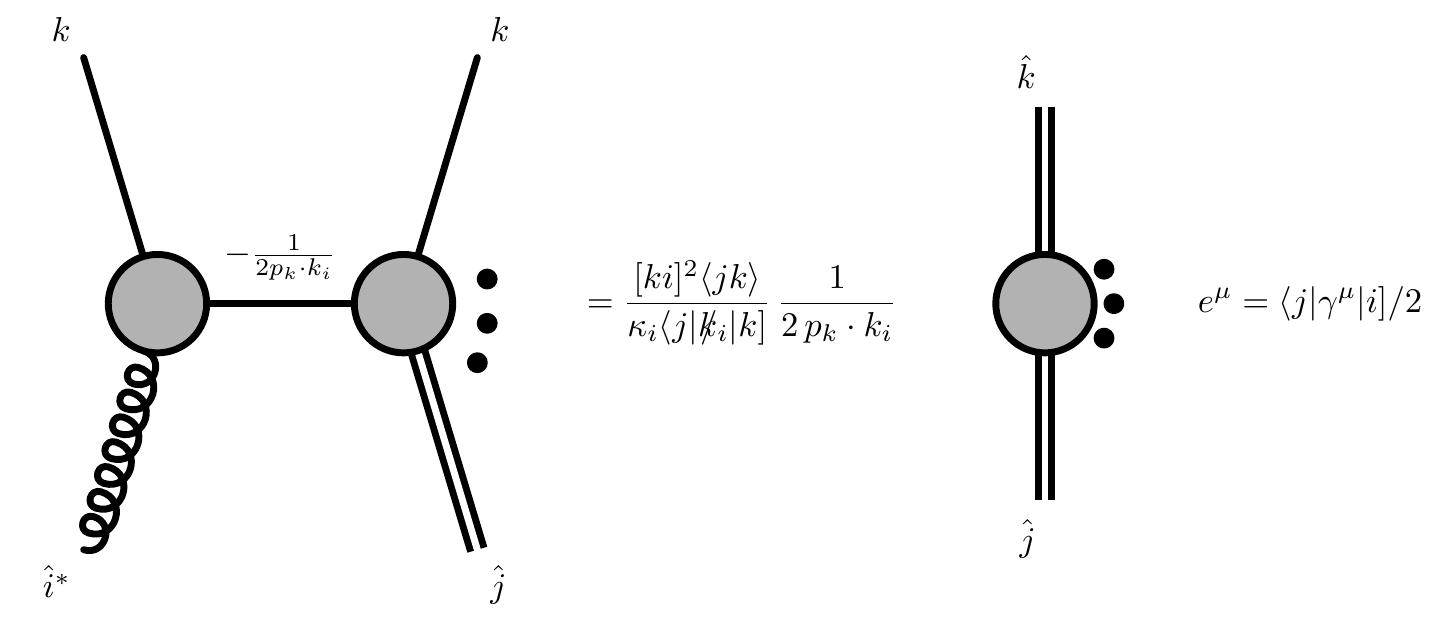}{60}{10}
\nn \\
&&
z = - \frac{2\,p_k\cdot k_i}{\SSS{ik}\AA{kj}} \, ,
\quad
\hat{\kapp}_k = \frac{\AL{j} \slashk_k + \slashk_i \SR{k}}{\AA{jk}} \, , 
\quad
\hat{\kstr}_k = \frac{\AL{k} \slashk_k + \slashk_i \SR{i}}{\SSS{ki}} \, ,
\nn \\
&&
\hat{\kapp}_j = \kapp_j - z\,\SSS{ji}  \, \lor \,  \SR{\hat{j}} = \SR{j} + z\, \SR{i}
\nn
\eea

\bea
&&
\graph{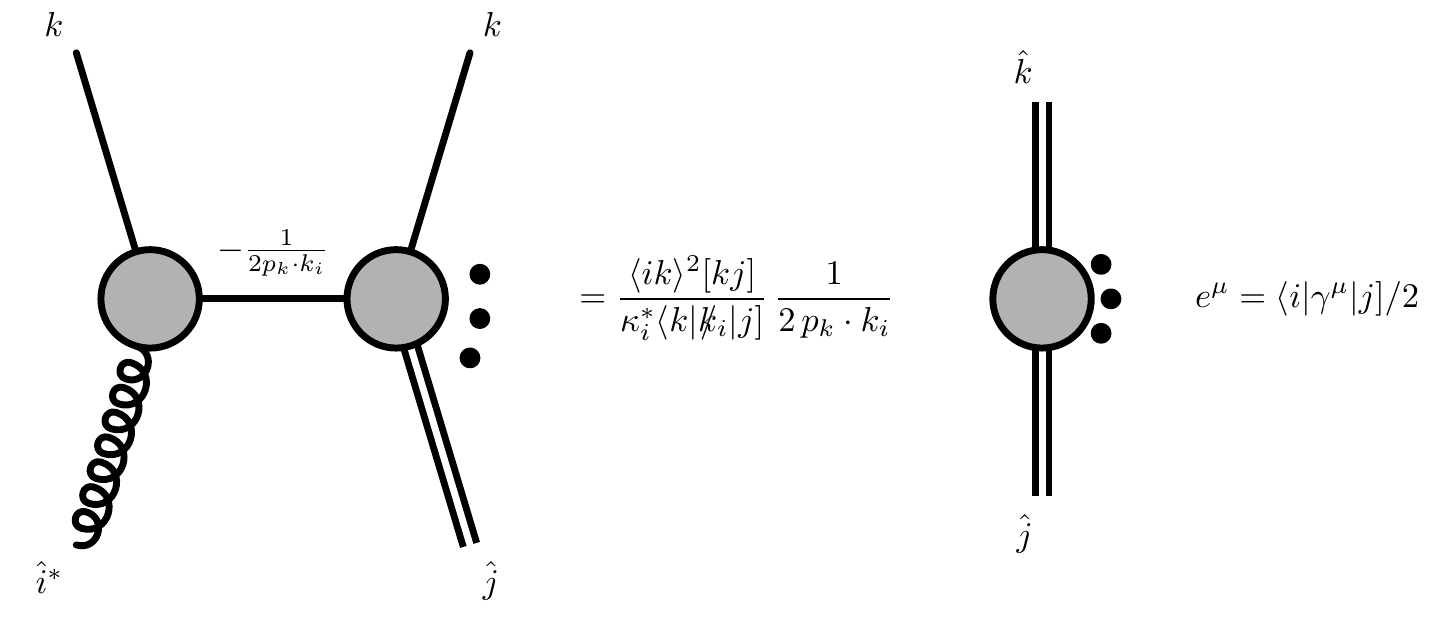}{60}{10}
\nn \\
&&
z = - \frac{2\,p_k\cdot k_i}{\AA{ik}\SSS{kj}} \, ,
\quad
\hat{\kapp}_k = \frac{\AL{i} \slashk_k + \slashk_i \SR{k}}{\AA{ik}} \, , 
\quad
\hat{\kstr}_k = \frac{\AL{k} \slashk_k + \slashk_i \SR{j}}{\SSS{kj}} \, ,
\nn \\
&&
\hat{\kstr}_j = \kstr_j + z\,\AA{ij}  \, \lor \,  \AR{\hat{j}} = \AR{j} - z\, \AR{i}
\nn
\eea
The bottom line is that, in these cases, in the $(n-2)$-point amplitude in the $B$-residue
there is one more shifted leg, according to the prescriptions reported above.   
As the situation with the 2-point amplitudes is rather specific, 
in order to indicate them during the derivations reported in the following we use the symbol $\Amp_2(\hat{K}^*,\hat{j})$.

Finally, $\mathrm{C}^{s}$ and $\mathrm{D}^{s}$ denote the same kind of residue, which appears respectively when the shifted $i$-th or $j$-th particle is off-shell.
They are due to the vanishing shifted momentum in the propagator of the off-shell particle: $k_i^2(z) = 0 \quad \text{or} \quad k_j^2(z)=0$.
These contributions are given by an amplitude with equal number of legs evaluated with the $i$-th or $j$-th leg on-shell but with momentum spinors
shifted according Tables \ref{CDg} and \ref{CDf} times proper prefactors. 
For gluons the prefactors are shown in the figure, for fermions the only difference is $x \rightarrow \sqrt{x}$. 
Let us recall that it was explicitly proven in \cite{vanHameren:2015bba} that, if the shift vector is chosen in such a way as to match the original LW
prescription in the on-shell limit, then the $C^f$ and $D^f$ terms vanish identically. This was enough there, where only one particle was off-shell;
now in principle both shifted particles (gluon and fermion) can be off-shell, so the statement is better recast in a slightly different way: 
the $C^f$ and $D^f$ residues are zero if the fermion spinor in the shift vector is the same as the one used in the diagrammatic construction of the amplitude. 
With our conventions (see also Eq. (\ref{Proper_spinors_1}) ), the correspondence is 
\beq \qb^+ \Leftrightarrow \SR{\qb}\, , \quad q^- \Leftrightarrow  \AL{q}\, , \quad \qb^- \Leftrightarrow \AR{\qb}\, , \quad q^+ \Leftrightarrow  \SL{q} \, .
\label{Proper_spinors_2}
\eeq
This statement can be checked to be equivalent to the previous one through the explicit analysis presented in \cite{vanHameren:2015bba}. 
As we will see first in the section dedicated the to 3-point amplitudes, also if both fermions are off-shell it is necessary to completely distinguish
the two spinor choices. This statement will also be explicitly illustrated with examples. This was not necessary with only one off-shell particle,
because in that case the spinor of the off-shell fermion is completely determined by the spinor used for the on-shell one, 
whose choice is in turn dictated by (\ref{Proper_spinors_2}).

Before presenting explicit results, one might ask whether the BCFW algorithm is necessarily the most efficient one 
when it comes to calculate multi-leg tree level amplitudes. In particular, it is known from \cite{Dinsdale:2006sq,Badger:2012uz} that,
at least for all-gluon amplitudes, it numerically outperforms the Berends-Giele off-shell recursion \cite{Berends:1987me} only for multiplicities $n < 9$. 
Besides, BCFW can be made more efficient by the use of hard-wired lower-point amplitudes, 
so that the recursion terminates with these instead of the 3-point amplitudes. 
There is no special reason to expect that fermion pairs should change much.
As far as the sole computation of the amplitudes is concerned, 
such a limiting multiplicity would be relevant for processes with $7$ particles final state or, maybe more realistically for the time being, for NNLO corrections
to $5$-jet production, which is the present benchmark multiplicity for NLO calculations achieved by the BLACK-HAT and NJET 
collaborations \cite{Bern:2011pa,Badger:2013yda}, if one assumes that the hurdles to 2-loop computations with final-state multiplicities 
higher than $2$ is to be overcome any time soon. 
Therefore, we can safely state that BCFW bears a twofold advantage: it employs, at every step, even in this modified version,
only gauge-invariant quantities and is numerically competitive with other methods.

\begin{table}
$$
\begin{array}{|c|c|}\hline
&    e^\mu = \AL{i}\g^\mu \SR{j} / 2
\\ 
\hline
&  (h_i, h_j) = (+,* \vee +) 
\\
\mathrm{C}^g 
& \SR{\hat{K}_i} = \sqrt{x_i}\, \SR{i} \, , \,\,\, \AR{\hat{K}_i} = \frac{\slashk_i\SR{j}}{\sqrt{x_i}\SSS{ij}} 
\\ 
& \kstrhat_j =   \frac{\AL{j} \slashk_i + \slashk_j \SR{i}}{\SSS{ji}}  \,\,\, \vee \,\,\, \AR{\hat{j}} = \frac{\left( \slashk_i + \slashp_j \right)\SR{i}}{\SSS{ji}} 
\\
\hline
&  (h_i, h_j) = (*\vee -,-) 
\\
\mathrm{D}^g 
& \AR{\hat{K}_j} = \sqrt{x_j}\, \AR{j} \, , \,\,\, \SR{\hat{K}_j} = \frac{\slashk_j\AR{i}}{\sqrt{x_j}\AA{ji}}
\\ 
& \kapphat_i =  \frac{\AL{j} \slashk_j + \slashk_i \SR{i}}{\AA{ij}}  \,\,\, \vee \,\,\, \SR{\hat{i}} = \frac{\left( \slashk_j + \slashp_i \right)\AR{i}}{\AA{ij}}
\\
\hline
\end{array}
$$
\caption{
Shifted quantities needed for the evaluation of $\mathrm{C}^g$ and $\mathrm{D}^g$ residues.
An obviously similar table holds for $ e^\mu = \AL{j}\g^\mu \SR{i} / 2  $ }
\label{CDg}
\end{table} 
\begin{table}
$$
\begin{array}{|c|c|c|}\hline
&   & \text{Admitted $(!)$ or not ($ \times $) on-shell }
\\ 
\hline
\mathrm{C}^f \Leftrightarrow e^\mu = \AL{\qb}\g^\mu \SR{g}/2
& \SR{\hat{K}_{\qb}} = \sqrt{x_{\qb}}\, \SR{\qb} \, , \,\,\, \AR{\hat{K}_{\qb}} = \frac{\slashk_{\qb} \SR{g}}{\sqrt{x_{\qb}} \SSS{\qb g}}  
&  (h_{\qb}, h_g) = (-,* \vee + ) \quad !
\\ 
& \kstrhat_g =   \frac{\AL{g} \slashk_{\qb} + \slashk_g \SR{\qb}}{\SSS{g \qb}}  \,\,\, \vee \,\,\, \AR{\hat{g}} = \frac{\left( \slashk_{\qb} + \slashp_g \right)\SR{{\qb}}}{\SSS{g \qb}} 
& (h_{\qb}, h_g) = (+,* \vee +) \quad \times
\\
\hline
\mathrm{D}^f \Leftrightarrow e^\mu = \AL{g}\g^\mu \SR{\qb}/2
& \AR{\hat{K}_{\qb}} = \sqrt{x_{\qb}}\, \AR{\qb} \, , \,\,\, \SR{\hat{K}_{\qb}} = \frac{\slashk_{\qb} \AR{g}}{\sqrt{x_{\bar{q}}}\AA{\qb g}}
& (h_g, h_{\qb}) = (* \vee -,+) \quad !
\\ 
& \kapphat_g =  \frac{\AL{\qb} \slashk_{\qb} + \slashk_g \SR{g}}{\AA{g \qb}}  \,\,\, \vee \,\,\, \SR{\hat{g}} = \frac{\left( \slashk_{\qb} + \slashp_g \right)\AR{g}}{\AA{g \qb}}
&  (h_g, h_{\qb}) = (* \vee -,-) \quad \times
\\
\hline
\end{array}
$$
\caption{
Shifted quantities needed for the evaluation of $\mathrm{C}^f$ and $\mathrm{D}^f$ residues in the case in which the antiquark is shifted.
A completely similar table holds for the case in which $q$ is shifted $( \bar{q} \rightarrow q)$.
}
\label{CDf}
\end{table} 
%

\section{The 3-point amplitudes}\label{3_Point_Amps}%

All the 3-point amplitudes with 1 off-shell leg have been provided in the first appendix of \cite{vanHameren:2015bba}.
There we showed their explicit diagrammatic construction, 
but also stressed the complete independence of the BCFW approach from any Feynman-diagram calculation; 
indeed, we also showed how off-shell 3-point amplitudes can be inferred via the BCFW procedure 
from their on-shell counterparts, which are in turn determined solely by symmetry requirements.
The same holds for this case, with the difference that the different kinematics
implies that we must take two residues into account, one for each off-shell leg.
This means that each of the amplitudes below is the sum of two terms, obtained by going on-shell with respect to both shifted off-shell legs in turn.
We do not provide any diagrammatic construction here but we just list the complete set of results.

\subsection{Two off-shell gluons}

The $3$ gluon amplitudes with two off-shell legs were given for 
the first time in \cite{vanHameren:2014iua} and are
\bea
\Amp(1^+,2^*,3^*) &=& \frac{1}{\kstr_2\kstr_3}\, \frac{\AA{23}^3}{\AA{31}\AA{12}} \, ,
\nn \\
\Amp(1^-,2^*,3^*) &=& \frac{1}{\kapp_2\kapp_3}\, \frac{\SSS{32}^3}{\SSS{21}\SSS{13}} \, .
\eea
%

\subsection{Off-shell gluon and off-shell fermion}

Concerning amplitudes with one off-shell gluon and one off-shell fermion, 
simple calculations give the results
\bea
\Amp(1^*,\qb^*,q^+) &=& \frac{1}{\kstr_1\,\kstr_{\qb}} \, \frac{\AA{1\qb}^3 \AA{1q}}{\AA{1\qb}\AA{\qb q}\AA{q 1}} \, ,
\nn \\
\Amp(1^*,\qb^*,q^-) &=& \frac{1}{\kapp_1\,\kapp_{\qb}} \, \frac{\SSS{1\qb}^3 \SSS{1q}}{\SSS{1q}\SSS{q \qb}\SSS{\qb 1}} \, ,
\nn \\
\Amp(1^*,\qb^+,q^*) &=& \frac{1}{\kstr_1\,\kstr_{q}} \, \frac{\AA{1 q}^3 \AA{1\qb}}{\AA{1\qb}\AA{\qb q}\AA{q 1}} \, ,
\nn \\
\Amp(1^*,\qb^-,q^*) &=& \frac{1}{\kapp_1\,\kapp_{q}} \, \frac{\SSS{1 q}^3 \SSS{1\qb}}{\SSS{1q}\SSS{q \qb}\SSS{\qb 1}} \, .
\label{offgq}
\eea

We explicitly show how to derive the first of (\ref{offgq}) through BCFW.
Let $e^\mu = \AL{1}\g^\mu\SR{\qb}/2$ be the shift vector. 
As $\qb$ cannot have but a negative helicity, such a shift vector does not respect the LW prescription.
Despite the fact that for an off-shell fermion it is still viable, the residue associated with the 
fermion external momentum going on-shell is not vanishing anymore.

From Tables \ref{CDg} and \ref{CDf} and taking prefactors properly into account, we see that
\bea
\Amp(1^*,\qb^*,q^+)  
&=& 
\frac{1}{x_g \kapp_g}\, \Amp(\hat{1}^+,\hat{\qb}^*,q^+) + \frac{1}{\sqrt{x_{\qb}\,\kstr_{\qb}}}\, \Amp(\hat{1}^*,\hat{\qb}^-,q^+) 
\nn \\
&=& 
0 + \frac{1}{\sqrt{x_{\qb}\,\kstr_{\qb}}}\, \frac{1}{\hat{\kstr_1}}\, \frac{\AA{\hat{K}_1 \hat{K}_{\qb}}^3\AA{\hat{K}_1q}}{\AA{\hat{K}_1\hat{K}_{\qb}} \AA{\hat{K}_{\qb} q} \AA{q\hat{K}_1}} \, ,
\eea
where the right side follows from the results derived in \cite{vanHameren:2015bba}.
Replacing the hatted spinors according to the results for the fermion $D$-pole in the same tables, we finally get the first line of (\ref{offgq}).

\subsection{Two off-shell fermions}

This is the most complicated case and it deserves further clarifications than the previous two.
First we report the results and then we comment on the meaning of the notation.
\bea
\Amp(1^+,\qb^{*+},q^{*-}) &=& \frac{1}{\kstr_{q}}\, \frac{\AA{q\qb}^3}{\AA{1\qb}\AA{\qb q}\AA{q1}}\, ,
\nn \\ 
\Amp(1^+,\qb^{*-},q^{*+}) &=& \frac{1}{\kstr_{\qb}}\, \frac{\AA{q\qb}^3}{\AA{1\qb}\AA{\qb q}\AA{q1}}\, ,
\nn \\
\Amp(1^-,\qb^{*+},q^{*-}) &=& \frac{1}{\kapp_{\qb}}\, \frac{\SSS{q\qb}^3}{\SSS{1q} \SSS{q\qb}\SSS{\qb 1}}\, ,
\nn \\ 
\Amp(1^-,\qb^{*-},q^{*+}) &=& \frac{1}{\kapp_{q}}\, \frac{\SSS{q\qb}^3}{\SSS{1q} \SSS{q\qb}\SSS{\qb 1}}\, .
\label{3ferfer}
\eea
The superscripts accompanying the off-shell fermions must be explained.
A fermion pair in an amplitude with all outgoing or all incoming particles must feature
opposite helicities and thus there are only two possibilities, namely
\begin{equation}
\Amp(\qb^+,q^-,g_1,\dots, g_n) =  \AL{q} \dots \SR{\qb} \, , \quad \Amp(\qb^-,q^+,g_1,\dots, g_n) =  \SL{q} \dots \AR{\qb} \, ,
\label{OrOr}
\end{equation}
where the spinors assignment is uniquely determined by the fermion helicity.
If only one of the fermions in the pair is off-shell, then its helicity is uniquely 
fixed by its partner's, which in turn determines the corresponding spinors, as in (\ref{OrOr}).
But if both fermions in the pair are off-shell, then one has to consider both options,
\begin{equation}
\Amp(\qb^{*+},q^{*-},g_1,\dots, g_n) =  \AL{q} \dots \SR{\qb} \, , \quad \Amp(\qb^{*-},q^{*+},g_1,\dots, g_n) =  \SL{q} \dots \AR{\qb} \, .
\label{AndAnd}
\end{equation}
The meaning of the superscripts should now be clear: for each off-shell pair 
and for every possible configuration of the other particles, 
there are two different off-shell amplitudes. 
These amplitudes are distinguished by the two possible different assignments 
of the fermion helicities.

In Appendix \ref{Calculus} we derive the first amplitude in two ways, first by explicit diagrammatic computation and then via BCFW recursion,
in order to fully illustrate the point for the interested reader.

\section{The 4-point amplitudes}\label{4_Point_Amps}

In this section we report the new computational results:
we present the analytical expressions for the four-point amplitudes with two off-shell particles.
In the cases with at least one external gluon, we employed the BCFW recursion and cross-checked
with the numerical results obtained with the help of {\sc AVHLIB}~\cite{Bury:2015dla}, 
which constitutes by itself a very non trivial cross-check.
All expressions for the amplitudes, except the four-fermion cases, have been
implemented for numerical evaluation in {\sc amp4hef}
\footnote{\tt https://bitbucket.org/hameren/amp4hef}.
Finally, all the fully squared matrix elements, averaged (summed) over initial (final) state colors and spins, 
were successfully cross-checked with the results reported in~\cite{Nefedov:2013ywa}. 

\subsection{Amplitudes with 4 gluons and no fermions}

These amplitudes were essentially all given already in \cite{vanHameren:2014iua}.
They are 
\begin{eqnarray}
%
&&
\left\{ \begin{array}{c}
\Amp(1^*,2^\pm,3^\pm,4^*) \\
\Amp(1^*,2^\pm,3^*,4^\pm) \\
\Amp(1^*,2^\pm,3^\mp,4^*) \\
\Amp(1^*,2^\pm,3^*,4^\mp) \\
%
\end{array} \right.
\nn
\end{eqnarray}
The first four amplitudes in the first two lines are Maximally Helicity Violating (MHV)
and are given by the general formulas (\ref{MHV_gluons}).

Concerning the other four amplitudes, the following two are calculated in \cite{vanHameren:2014iua}:
\bea
\Amp(1^*,2^-,3^*,4^+)
&=&
\frac{1}{\kstr_1\kapp_3}\, \frac{\AA{12}^3\SSS{43}^3}{ \AL{2} \slashk_3 \SR{4}\, \AL{1} \slashk_3 + \slashp_4 \SR{3}\, \left( k_3+p_4\right)^2} 
\nn \\
&+&
\frac{1}{\kapp_1\kstr_3}\, \frac{\AA{23}^3\SSS{14}^3}{ \AL{2} \slashk_1 \SR{4}\, \AL{3} \slashk_1 + \slashp_4 \SR{1}\, \left( k_1+p_4\right)^2} \, ,
\nn \\
\Amp(1^*,2^+,3^-,4^*)
&=&
- \frac{1}{\kstr_1\kapp_4}\, \frac{\AL{1} \slashk_4 + \slashp_3  \SR{4}^4 }{\AL{2} \slashk_1  \SR{4}  \, \AL{1} \slashk_4  \SR{3 } \, \AA{12} \, \SSS{43}\, (k_4+p_3)^2}
\nn \\
&+&
\frac{1}{\kapp_1}\, \frac{\AA{34}^3\, \SSS{14}^3 }{\AL{4} \slashk_1 + \slashk_4 \SR{1} \, \AL{2} \slashk_1 \SR{4} \, \AL{4} \slashk_1 \SR{4} \, \AA{23} } 
\nn \\
&+&
\frac{1}{\kstr_4}\, \frac{\AA{14}^3\, \SSS{21}^3 }{\AL{4} \slashk_1 + \slashk_4 \SR{1} \, \AL{1} \slashk_4 \SR{3} \, \AL{1} \slashk_4 \SR{1} \, \SSS{32} } \, .
\label{Andreas_didit}
\eea
From cyclic invariance it is apparent that $\Amp(1^*,2^+,3^*,4^-)$ can be obtained 
from the first amplitude in (\ref{Andreas_didit}) just by the relabelings $(1 \leftrightarrow 3, 2 \leftrightarrow 4)$. 
In order to get $\Amp(1^*,2^-,3^+,4^*)$ from the second, instead, it is necessary to assume that the color-order
reversed relation holds,
\beq
\Amp(1,2,\dots,n) = (-1)^n\, \Amp(n,n-1,\dots,2,1) \, ,
\label{CORR}
\eeq
before performing the relabelings $(1 \leftrightarrow 4, 2 \leftrightarrow 3)$.
We checked explicitly that this is actually the case for this amplitude and in Appendix
\ref{Feng} we recall the short and beautiful proof of (\ref{CORR}) presented in
\cite{Feng:2010my} and show that it carries over directly to our amplitudes with off-shell legs as well.

\subsection{Amplitudes with 2 gluons and 1 fermion pair}

\subsubsection{2 off-shell gluons}\label{gstargstar}

There are two such amplitudes, namely $\Amp(2^*,\qb^\pm,q^\mp,1^*)$.
So we compute the first one explicitly in Appendix \ref{Calculus}, the other one being analogous.
Here we just report the final results, 
\bea 
\Amp(2^*,\qb^+,q^-,1^*) 
&=& 
\frac{1}{\kapp_1} \, \frac{\AA{2q}^3 \, \SSS{12}^3 }{ \AA{\qb q} \, \AL{2} \slashk_2 + \slashk_1 \SR{1} \, \AL{q} \slashk_1 \SR{2} \, \AL{2} \slashk_1 \SR{2} }
\nn \\
&+&
\frac{1}{\kstr_2} \, \frac{\AA{21}^3 \, \SSS{1\qb}^3}{ \SSS{\qb q} \, \AL{2} \slashk_2 + \slashk_1 \SR{1} \, \AL{1} \slashk_2 \SR{\qb} \, \AL{1} \slashk_2 \SR{1} }
\nn \\
&-&
\frac{1}{\kstr_1\,\kapp_2} \, \frac{1}{(k_2+p_{\qb})^2} \, 
\frac{ \SSS{2\qb}^2 \, \AA{1q}^2 \, \AL{1} \slashk_2+\slashp_{\qb} \SR{2} }{\AL{q} \slashk_2 \SR{\qb} \, \AL{1} \slashk_2 + \slashp_{\qb} \SR{2} + (k_2+p_{\qb})^2\, \SSS{\qb2} \, \AA{q1}} \, .
\label{2offg_1}
\\
%
\Amp(2^*,\qb^-,q^+,1^*) 
&=&
\frac{1}{\kstr_1} \, \frac{ \SSS{2q}^3 \, \AA{12}^3  }{\SSS{q\qb} \, \AL{1} \slashk_1 + \slashk_2 \SR{2} \, \AL{2} \slashk_1 \SR{q} \, \AL{2} \slashk_1 \SR{2} }
\nn \\
&+&
\frac{1}{\kapp_2} \, \frac{\SSS{21}^3 \, \AA{1\qb}^3 }{\AA{q\qb} \, \AL{1} \slashk_1 + \slashk_2 \SR{2} \, \AL{\qb} \slashk_2 \SR{1}\, \AL{1} \slashk_2 \SR{1}}
\nn \\
&-&
\frac{1}{\kapp_1 \, \kstr_2} \, \frac{1}{(k_2+p_{\qb})^2} \, 
\frac{ \AA{2\qb}^2 \, \SSS{1q}^2 \, \AL{2} \slashk_2+\slashp_{\qb} \SR{1} }{\AL{\qb} \slashk_2 \SR{q} \, \AL{2} \slashk_2 + \slashp_{\qb} \SR{1} + (k_2+p_{\qb})^2\, \AA{\qb2}\,\SSS{q1}}
\, .
\label{2offg_2}
\eea
As expected, (\ref{2offg_1}) and (\ref{2offg_2}) are manifestly each other's parity conjugated amplitudes.
They were obtained with shift vectors $e^\mu = \AL{1}\gamma^\mu\SR{2}/2$ and $e^\mu = \AL{2}\gamma^\mu\SR{1}/2$ respectively.

\subsubsection{1 off-shell gluon and 1 off-shell fermion}

This is the most complicated case, since these amplitude also depend on the 
position of the off-shell gluon relative to the fermion pair, a situation already met
in \cite{vanHameren:2015bba}.

There are $8$ independent such amplitudes, 
the other 8 being parity conjugated and thus given simply by their adjoints.
The independent ones can conveniently be grouped into MHV and NMHV amplitudes, i.e.
\begin{eqnarray}
\textrm{MHV} \quad
&&
\left\{ \begin{array}{c}
\Amp(2^*,\qb^+,q^*,1^+) \\
\Amp(2^*,\qb^*,q^+,1^+) \\
\Amp(2^+,\qb^+,q^*,1^*) \\
\Amp(2^+,\qb^*,q^+,1^*) \\
%
\end{array} \right.
\nn
%
\\
%
%
\textrm{NMHV} \quad
&&
\left\{ \begin{array}{c}
\Amp(2^*,\qb^+,q^*,1^-) \\
\Amp(2^*,\qb^*,q^+,1^-) \\
\Amp(2^-,\qb^+,q^*,1^*) \\
\Amp(2^-,\qb^*,q^+,1^*) \\
%
\end{array} \right.
\nn
\end{eqnarray}
The general formulas for MHV amplitudes were already given in section \ref{MHV}.

As for the NMHV ones, we have to compute 4 of them.
We explicitly show the calculation of the first one in Appendix \ref{Calculus}, 
providing only the results for the remaining three
and for all the adjoint amplitudes. 

Here are the final expressions for all the 8 NMHV amplitudes
\bea
\Amp(2^*,\qb^+,q^*,1^-) 
&=&
\frac{1}{\kstr_2} \, \frac{\SSS{q\qb}^2 \, \AA{2q}^2 }{ \SSS{q1} \, \AL{q} \slashk_2  \SR{1} \, \AL{q} \slashk_2 \SR{\qb} } 
\nn \\ 
&& \hspace{-12mm }
- \frac{1}{\kapp_2\,\kstr_q}\, \left(
\frac{1}{(k_2 + p_{\qb})^2} \, \frac{ \SSS{2\qb}^2 \, \AA{1q}^2 }{ \AL{q} \slashk_2 \SR{\qb} } - 
\frac{1}{(p_1 + k_2)^2} \,\frac{ \AL{q} \slashk_2 + \slashp_1 \SR{2}^3 }{ \SSS{12} \, \AA{q\qb}  \, \AL{q} \slashk_2 \SR{1} }
\right) \, ,
\nn \\
\Amp(2^*,\qb^-,q^*,1^+) 
&=&
\frac{1}{\kapp_2} \, \frac{\AA{q\qb}^2 \, \SSS{2q}^2 }{ \AA{1q} \, \AL{1} \slashk_2  \SR{q} \, \AL{\qb} \slashk_2 \SR{q} } 
\nn \\ 
&& \hspace{-12mm }
- \frac{1}{\kstr_2\,\kapp_q}\, \left(
\frac{1}{(k_2 + p_{\qb})^2} \, \frac{ \AA{2\qb}^2 \, \SSS{1q}^2 }{ \AL{\qb} \slashk_2 \SR{q} } - 
\frac{1}{(p_1 + k_2)^2} \,\frac{ \AL{2} \slashk_2 + \slashp_1 \SR{q}^3 }{ \AA{12} \, \SSS{q\qb}  \, \AL{1} \slashk_2 \SR{q} }
\right) \, .
\nn \\
\Amp(2^-,\qb^*,q^+,1^*) 
&=&
\frac{1}{\kstr_1} \, \frac{\SSS{q\qb}^2 \, \AA{1\qb}^2 }{ \SSS{2\qb} \, \AL{\qb} \slashk_1  \SR{2} \, \AL{\qb} \slashk_1 \SR{q} } 
\nn \\ 
&& \hspace{-12mm }
- \frac{1}{\kapp_1\,\kstr_{\qb}}\, \left(
\frac{1}{(k_1 + p_q)^2} \, \frac{ \SSS{1q}^2 \, \AA{2\qb}^2 }{ \AL{\qb} \slashk_1 \SR{q} } - 
\frac{1}{(p_2 + k_1)^2} \,\frac{ \AL{\qb} \slashk_1 + \slashp_2 \SR{1}^3 }{ \SSS{12} \, \AA{q\qb}  \, \AL{\qb} \slashk_1 \SR{2} }
\right) \, ,
\nn \\
\Amp(2^+,\qb^*,q^-,1^*) 
&=&
\frac{1}{\kapp_1} \, \frac{\AA{q\qb}^2 \, \SSS{1\qb}^2 }{ \AA{\qb2} \, \AL{2} \slashk_1  \SR{\qb} \, \AL{q} \slashk_1 \SR{\qb} } 
\nn \\ 
&& \hspace{-12mm }
- \frac{1}{\kstr_1 \, \kapp_{\qb}}\, \left(
\frac{1}{(k_1 + p_q)^2} \, \frac{ \AA{1q}^2 \, \SSS{2\qb}^2 }{ \AL{q} \slashk_1 \SR{\qb} } - 
\frac{1}{(p_2 + k_1)^2} \,\frac{ \AL{1} \slashk_1 + \slashp_2 \SR{\qb}^3 }{ \AA{12} \, \SSS{q\qb}  \, \AL{2} \slashk_1 \SR{\qb} }
\right) \, ,
\nn \\
\Amp(2^*,\qb^*,q^+,1^-) 
&=&
\frac{1}{\kstr_2} \, \frac{\SSS{q\qb}^2 \, \AA{2\qb}^2 }{\SSS{1q} \, \AL{\qb} \slashk_2 \SR{1} \, \AL{\qb} \slashk_2 \SR{\qb} } +
\frac{1}{\kapp_2\,\kstr_{\qb}} \, \frac{1}{(p_1+k_2)^2} \, \frac{\AL{\qb} \slashp_1 + \slashk_2 \SR{2}^3}{\SSS{12} \, \AA{q\qb} \, \AL{\qb} \slashk_2 \SR{1}} \, ,
\nn \\
\Amp(2^*,\qb^*,q^-,1^+) 
&=&
\frac{1}{\kapp_2} \, \frac{\AA{q\qb}^2 \, \SSS{2\qb}^2 }{\AA{q1} \, \AL{1} \slashk_2 \SR{\qb} \, \AL{\qb} \slashk_2 \SR{\qb} } +
\frac{1}{\kstr_2\,\kapp_{\qb}} \, \frac{1}{(p_1+k_2)^2} \, \frac{\AL{2} \slashp_1 + \slashk_2 \SR{\qb}^3 }{\AA{12} \, \SSS{q\qb} \, \AL{1} \slashk_2 \SR{\qb}} \, ,
\nn \\
\Amp(2^-,\qb^+,q^*,1^*) 
&=&
\frac{1}{\kstr_1} \, \frac{ \SSS{q\qb}^2 \, \AA{1q}^2 }{ \SSS{2\qb} \, \AL{q} \slashk_1 \SR{2} \, \AL{q} \slashk_1 \SR{q} } + 
\frac{1}{\kapp_1\,\kstr_q} \, \frac{1}{(k_1+p_2)^2} \, \frac{\AL{q} \slashk_1+\slashp_2 \SR{1}^3}{\SSS{12} \, \AA{q\qb} \, \AL{q} \slashk_1 \SR{2}} \, ,
\nn \\
\Amp(2^+,\qb^-,q^*,1^*) 
&=&
\frac{1}{\kapp_1} \, \frac{ \AA{q\qb}^2 \, \SSS{1q}^2 }{ \AA{\qb2} \, \AL{2} \slashk_1 \SR{q} \, \AL{q} \slashk_1 \SR{q} } + 
\frac{1}{\kstr_1 \, \kapp_q} \, \frac{1}{(k_1+p_2)^2} \, \frac{ \AL{1} \slashk_1 + \slashp_2 \SR{q}^3 }{ \AA{12} \, \SSS{q\qb} \, \AL{2} \slashk_1 \SR{q} } \, .
\eea

\subsubsection{2 off-shell fermions}

The final set of amplitudes we can evaluate with BCFW recursion are the ones with two on-shell gluons
and one off-shell fermion pair. There are 8 of them, 4 of the kind $\Amp(2^\pm,\qb^*,q^*,1^\pm)$ and 
4 of the kind $\Amp(2^\pm,\qb^*,q^*,1^\mp)$, because the fermion pair, as detailed in section \ref{3_Point_Amps},
can come in two ways, depending on the way the spinors are assigned. \\
The formulas for the MHV amplitudes $\Amp(2^\pm,\qb^*,q^*,1^\pm)$ were given in section \ref{MHV}. \\
The $4$ non MHV amplitudes require some work, which is explicitly done for two of them which are non parity-conjugated
in the Appendix \ref{Calculus}.

The final explicit results are
\bea
\Amp(2^+,\qb^{*+},q^{*-},1^-) 
&=&
\frac{1}{\kappa_{\qb}\,\kstr_{q}} \, 
\frac{1}{(p_2+k_{\qb})^2} \, 
\frac{\SSS{2\qb}^2 \, \AA{1q}^2 \, \AL{1} \slashk_{\qb} \SR{2} }{ \AL{q} \slashp_2 + \slashk_{\qb} \SR{\qb} \, \AL{1} \slashk_{\qb} \SR{2} + (p_2+k_{\qb})^2 \, \AA{1q} \, \SSS{2\qb}}
\nn \\
&+& 
\frac{ \AA{1\qb}^4 \, \SSS{q\qb}^2 }{ \AA{12} \, \AA{2\qb} \, \AL{1} \slashk_{\qb} + \slashp_2 \SR{\qb} \, \AL{\qb} \, \slashp_1 + \slashp_2 \SR{q} \, \AL{\qb} \slashp_1 + \slashp_2 \SR{\qb} }
\nn \\
&+&
\frac{ \SSS{2q}^4 \, \AA{q\qb}^2 }{ \SSS{12} \, \SSS{1q} \, \AL{q} \slashk_q + \slashp_1 \SR{2} \, \AL{\qb} \slashp_1 + \slashp_2 \SR{q} \, \AL{q} \slashp_1 + \slashp_2 \SR{q} } \, ,
\nn \\
\Amp(2^-,\qb^{*-},q^{*+},1^+)
&=&
\frac{1}{\kstr_{\qb}\,\kapp_{q}} \, 
\frac{1}{(p_2+k_{\qb})^2} \, 
\frac{\AA{2\qb}^2 \, \SSS{1q}^2 \, \AL{2} \slashk_{\qb} \SR{1} }{ \AL{\qb} \slashp_2 + \slashk_{\qb} \SR{q} \, \AL{2} \slashk_{\qb} \SR{1} + (p_2+k_{\qb})^2 \, \SSS{1q} \, \AA{2\qb}}
\nn \\
&+& 
\frac{ \SSS{1\qb}^4 \, \AA{q\qb}^2 }{ \SSS{12} \, \SSS{2\qb} \, \AL{\qb} \slashk_{\qb} + \slashp_2 \SR{1} \, \AL{q} \, \slashp_1 + \slashp_2 \SR{\qb} \, \AL{\qb} \slashp_1 + \slashp_2 \SR{\qb} }
\nn \\
&+&
\frac{ \AA{2q}^4 \, \SSS{q\qb}^2 }{ \AA{12} \, \AA{1q} \, \AL{2} \slashk_q + \slashp_1 \SR{q} \, \AL{q} \slashp_1 + \slashp_2 \SR{\qb} \, \AL{q} \slashp_1 + \slashp_2 \SR{q} } \, .
\nn \\
\Amp(2^+,\qb^{*-},q^{*+},1^-) 
&=&
\frac{1}{\kapp_q} \, \frac{\AA{1\qb}^3 \, \SSS{q\qb}^2}{\AA{12} \, \AA{2\qb} \, \AL{\qb} \, \slashp_1+\slashp_2  \SR{q} \, \AL{\qb} \slashp_1+\slashp_2 \SR{\qb} } 
\nn \\
&+&
\frac{1}{\kstr_{\qb}} \, \frac{\SSS{2q}^3 \, \AA{q\qb}^2}{\SSS{12} \, \SSS{q1} \, \AL{\qb} \slashp_1+\slashp_2 \SR{q} \, \AL{q} \slashp_1+\slashp_2 \SR{q}} \, ,
\nn \\
\Amp(2^-,\qb^{*+},q^{*-},1^+) 
&=&
\frac{1}{\kstr_q} \, \frac{\SSS{\qb1}^3 \, \AA{q\qb}^2}{\SSS{12} \, \SSS{2\qb} \, \AL{q} \, \slashp_1+\slashp_2  \SR{\qb} \, \AL{\qb} \slashp_1+\slashp_2 \SR{\qb} } 
\nn \\
&+&
\frac{1}{\kapp_{\qb}} \, \frac{\AA{q2}^3 \, \SSS{q\qb}^2}{\AA{12} \, \AA{q1} \, \AL{q} \slashp_1+\slashp_2 \SR{\qb} \, \AL{q} \slashp_1+\slashp_2 \SR{q}} \, .
\eea
%

\subsection{Amplitudes with 4 fermions and no gluons}

\subsubsection{Amplitudes with 2 different-flavour fermion pairs}

In order to compute this amplitudes, we can no longer apply the BCFW recursion, as the condition
of having at least one external gauge boson is not fulfilled. So, as we want to provide a full set of 
4-point amplitudes for high-energy factorization, we have to compute them via Feynman diagrams expansion.

\begin{figure}[ht]
\begin{center}
\includegraphics[scale=0.80]{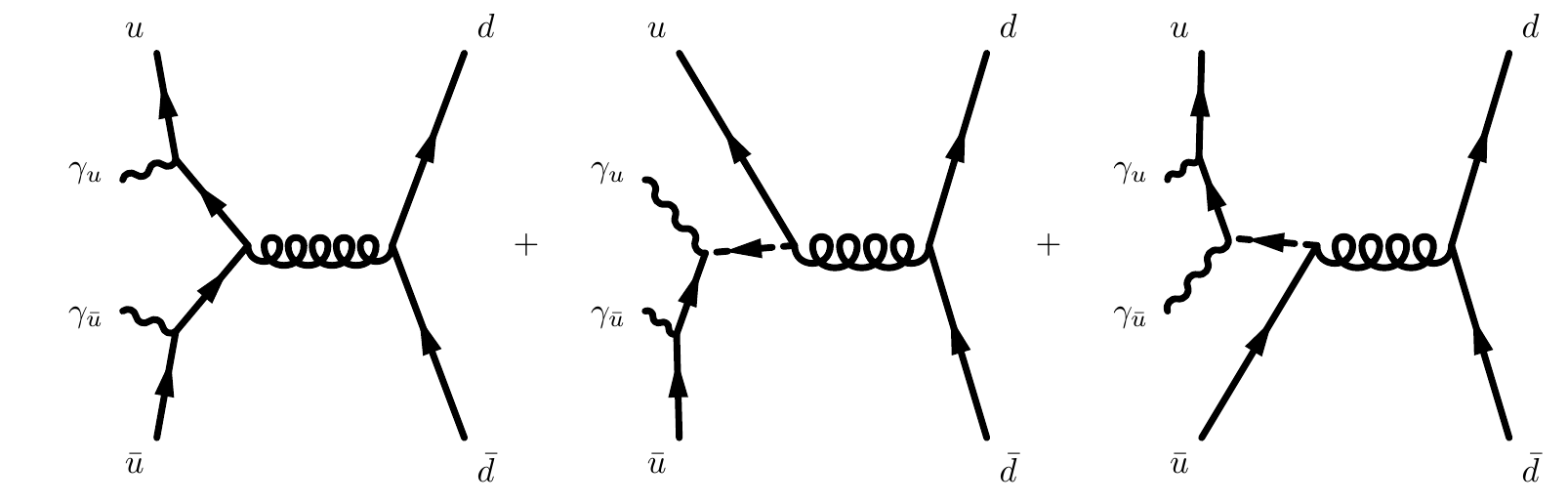}
\caption{The three contributions to $\Amp(\ub^{*+},u^{*-},\db^{+},d^{-})$ 
and $\Amp(\ub^{*+},u^{*-},\db^{-},d^{+})$}
\label{Fig_offubuondbd}
\end{center}
\end{figure}

The case with two quark pairs of different flavours is simpler and we will start with it.
Let us use the letters `u' and `d' for different flavours.  
Modulo parity conjugation, the independent amplitudes are
\bea
&&
\Amp(\ub^{*+},u^{*-},\db^{+},d^{-}) \, , \quad \Amp(\ub^{*+},u^{*-},\db^{-},d^{+})
\nn \\
&&
\Amp(\ub^{*+},u^{-},\db^{*+},d^{-}) \, , \quad \Amp(\ub^{*+},u^{-},\db^{*-},d^{+})
%
\eea
We start with the case with an off-shell quark-anti-quark pair of the same flavour, which is diagrammatically represented in
Fig. \ref{Fig_offubuondbd}

Using the Feynman rules, one can calculate quite easily that
\bea
\Amp(\ub^{*+},u^{*-},\db^{+},d^{-}) 
&=&
\frac{\SSS{\db \ub} \, \AA{du}}{ \kapp_{\ub} \, \kstr_{u}\, \SSS{d\db} \, \AA{d\db}} +
\frac{\SSS{\db u} \, \SSS{\ub u } \, \AA{du} }{ \kapp_{\ub} \, \SSS{d\db}\, \AA{d\db} \, \AL{u} \slashk_{\ub} \SR{u}} +
\frac{\SSS{\db \ub} \, \AA{\ub u} \, \AA{d \ub} }{ \kstr_{u} \, \SSS{d \db}\, \AA{d \db} \, \AL{\ub} \slashk_{u} \SR{\ub}}
\nn\\
\Amp(\ub^{*+},u^{*-},\db^{-},d^{+})
&=&
\frac{\SSS{d \ub} \, \AA{\db u}}{ \kapp_{\ub} \, \kstr_{u} \, \SSS{d\db} \, \AA{d\db}} +
\frac{\SSS{d u} \, \SSS{\ub u } \, \AA{\db u} }{ \kapp_{\ub} \, \SSS{d\db}\, \AA{d\db} \, \AL{u} \slashk_{\ub} \SR{u}} +
\frac{\SSS{d \ub} \, \AA{\ub u} \, \AA{\db \ub} }{ \kstr_{u} \, \SSS{d \db}\, \AA{d \db} \, \AL{\ub} \slashk_{u} \SR{\ub}}
\nn \\
\eea

Next is the configuration with two off-shell antiquarks, which is made by the four contributions depicted in Fig. \ref{Fig_offubuondbd}.
\begin{figure}[ht]
\begin{center}
\includegraphics[scale=0.80]{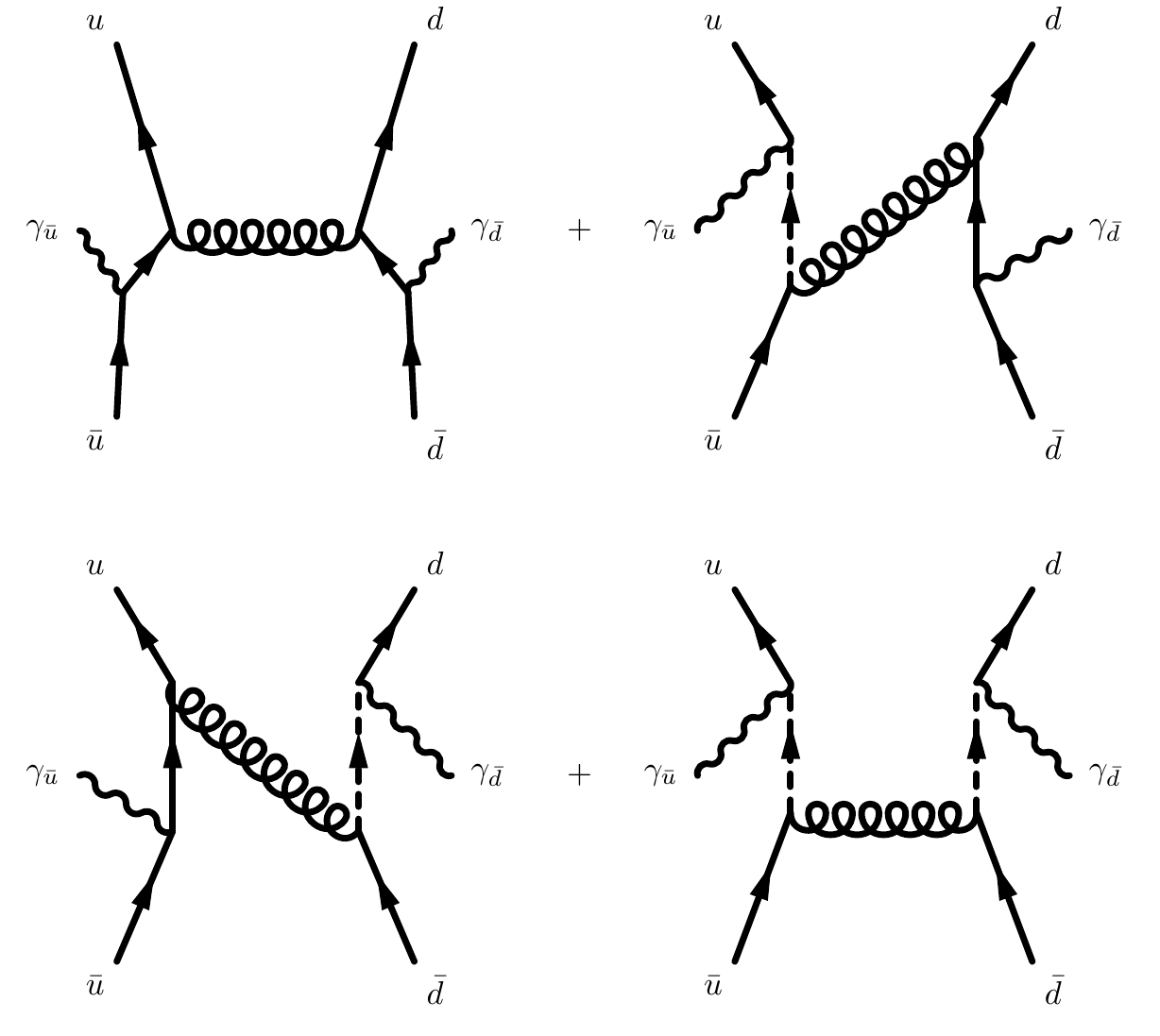}
\caption{The four contributions to $\Amp(\ub^{*+},u^{-},\db^{*+},d^{-})$ 
and $\Amp(\ub^{*+},u^{-},\db^{*-},d^{+})$}
\label{Fig_offubdbonud}
\end{center}
\end{figure}
Applying the aforementioned Feynman rules to the four contributions shown in Fig. \ref{Fig_offubdbonud}, we arrive at 
\bea
\Amp(\ub^{*+},u^{-},\db^{*+},d^{-}) 
&=&
\frac{	\SSS{\db \ub} \, \AA{d u}}{\kapp_{\ub} \, \kapp_{\db} \, \left( p_d + k_{\db} \right)^2} +
\frac{\AA{\db u} \, \SSS{\db \ub} }{\kapp_{\ub} \, \left( p_d+k_{\db} \right)^2 \, \SSS{d\db} } +
\frac{\SSS{\db \ub} \, \AA{d \ub}  }{\kapp_{\db} \, \left( p_d + k_{\db} \right)^2 \, \SSS{u \ub} }
\nn \\
&+&
\frac{ \SSS{\db \ub} \, \AA{\db\ub} }{\left( p_d + k_{\db} \right)^2 \, \SSS{ d\db} \, \SSS{u \ub}}
\nn \\
\Amp(\ub^{*+},u^{-},\db^{*-},d^{+})
&=&
\frac{	\SSS{d \ub} \, \AA{\db u}}{\kapp_{\ub} \, \kstr_{\db} \, \left( p_d + k_{\db} \right)^2} +
\frac{\AA{u \db} \, \SSS{\db \ub} }{\kapp_{\ub} \, \left( p_d+k_{\db} \right)^2 \, \AA{d\db} } +
\frac{ \SSS{d \ub} \, \AA{\db \ub} }{\kstr_{\db} \, \left( p_d + k_{\db} \right)^2 \, \SSS{u \ub} }
\nn \\
&+&
\frac{ \SSS{\db \ub} \, \AA{\ub\db} }{\left( p_d + k_{\db} \right)^2 \, \AA{ d\db} \, \SSS{u \ub}}
\eea
%

\subsubsection{Amplitudes with 2 equal-flavour fermion pairs}

We come to the last two sets of amplitudes, which we need to describe the processes 
$q^*\,q^* \rightarrow q q$ and  $\bar{q}^*\,q^* \rightarrow \bar{q} \, q$.

Concerning the first process, we have the four independent contributions
\bea
\Amp(u_1^{*+},\ub_1^-,\ub_2^-,u_2^{*+})  
&=&
\frac{\SSS{u_1 u_2}\, \AA{\ub_1 \ub_2} }{\kapp_{u_1}\, \kapp_{u_2}\,\left( k_{u_1} + p_{\ub_1} \right)^2 } +  
\frac{\AA{u_1 \ub_2} \, \SSS{u_1 u_2} }{\kapp_{u_2} \, \left( k_{u_1} + p_{\ub_1} \right)^2 \, \SSS{\ub_1 u_1}}
\nn \\
&+&
\frac{\AA{\ub_1 u_2} \, \SSS{u_1 u_2}}{\kapp_{u_1}\, \left( k_{u_1} + p_{\ub_1} \right)^2 \, \SSS{\ub_2 u_2}} +  
\frac{\AA{u_1 u_2} \, \SSS{u_1 u_2} }{ \left( k_{u_1} + p_{\ub_1} \right)^2 \, \SSS{\ub_1 u_1} \, \SSS{\ub_2 u_2} } \, ,
\nn \\
\Amp(u_1^{*+},\ub_1^-,\ub_2^+,u_2^{*-}) 
&=&
\frac{\SSS{u_1 \ub_2}\, \AA{\ub_1 u_2} }{\kapp_{u_1}\, \kstr_{u_2}\,\left( k_{u_1} + p_{\ub_1} \right)^2 } +
\frac{\AA{u_1 u_2} \, \SSS{u_1 \ub_2} }{\kstr_{u_2} \, \left( k_{u_1} + p_{\ub_1} \right)^2 \, \SSS{\ub_1 u_1}}
\nn \\
&-&
\frac{\AA{\ub_1 u_2} \, \SSS{u_1 u_2}}{\kapp_{u_1}\, \left( k_{u_1} + p_{\ub_1} \right)^2 \, \AA{\ub_2 u_2}} -
\frac{\AA{u_1 u_2} \, \SSS{u_1 u_2} }{ \left( k_{u_1} + p_{\ub_1} \right)^2 \, \SSS{\ub_1 u_1} \, \AA{\ub_2 u_2} } \, , 
\nn \\
\Amp(u_1^{*+},\ub_2^-,\ub_1^-,u_2^{*+}) 
&=&
- \frac{\SSS{u_1 u_2}\, \AA{\ub_1 \ub_2} }{\kapp_{u_1}\, \kapp_{u_2}\,\left( k_{u_1} + p_{\ub_2} \right)^2 } +  
\frac{\AA{\ub_1 u_1} \, \SSS{u_1 u_2} }{\kapp_{u_2} \, \left( k_{u_1} + p_{\ub_2} \right)^2 \, \SSS{ u_1 \ub_2}}
\nn \\
&+&
\frac{\AA{\ub_2 u_2 } \, \SSS{u_1 u_2}}{\kapp_{u_1}\, \left( k_{u_1} + p_{\ub_2} \right)^2 \, \SSS{\ub_1 u_2 }} -  
\frac{\AA{u_1 u_2} \, \SSS{u_1 u_2} }{ \left( k_{u_1} + p_{\ub_2} \right)^2 \, \SSS{u_1 \ub_2} \, \SSS{\ub_1 u_2}} \, , 
\nn \\
\Amp(u_1^{*+},\ub_2^-,\ub_1^+,u_2^{*-}) 
&=&
- \frac{\SSS{\ub_1 u_1}\, \AA{\ub_2 u_2} }{\kapp_{u_1}\, \kstr_{u_2}\,\left( k_{u_1} + p_{\ub_2} \right)^2 } +
\frac{\AA{u_1 u_2} \,  \SSS{\ub_1 u_1 } }{\kstr_{u_2} \, \left( k_{u_1} + p_{\ub_2} \right)^2 \, \SSS{u_1 \ub_2}}
\nn \\
&-&
\frac{\AA{\ub_2 u_2 } \, \SSS{u_1 u_2}}{\kapp_{u_1}\, \left( k_{u_1} + p_{\ub_2} \right)^2 \, \AA{\ub_1 u_2}} +
\frac{\AA{u_1 u_2} \, \SSS{u_1 u_2} }{ \left( k_{u_1} + p_{\ub_2} \right)^2 \, \SSS{u_1 \ub_2} \, \AA{\ub_1 u_2 }}  \, .
\eea
As we are working in an all-incoming setup, the letter $u$ denotes the off-shell quarks and
$\ub$ stands for the on-shell antiquarks, which become outgoing quarks when switching their momenta signs. 
The particles with the label $1$ together make the $t$-channel.

For the second process instead, the four independent contributions are
\bea 
\Amp( \bar{u}_1^{*+},u_1^{*-},\ub_2^+,u_2^-) 
&=&
\frac{\SSS{\ub_1 \ub_2} \AA{u_1 u_2} }{\kapp_{\ub_1}\, \kstr_{u_1} \, \SSS{\ub_2 u_2}\, \AA{\ub_2 u_2} } + 
\frac{\AA{\ub_1 u_1} \, \AA{ \ub_1 u_2} \, \SSS{\ub_1 \ub_2} }{\kstr_{u_1} \, \AL{\ub_1}\slashk_{u_1} \SR{\ub_1} \, \AA{\ub_2 u_2}\, \SSS{\ub_2 u_2}\, } 
\nn \\
&+&
\frac{\AA{u_1 u_2} \, \SSS{\ub_1 u_1 } \, \SSS{u_1 \ub_2} }{\kapp_{\ub_1} \, \AL{u_1}\slashk_{\ub_1} \SR{u_1} \, \AA{\ub_2 u_2 }\, \SSS{\ub_2 u_2} } \, , 
\nn \\
\Amp( \bar{u}^{*+}_1,u_1^{*-},\ub_2^-,u_2^+) 
&=&
\frac{\SSS{\ub_1 u_2} \AA{u_1 \ub_2} }{\kapp_{\ub_1}\, \kstr_{u_1} \, \SSS{\ub_2 u_2}\, \AA{\ub_2 u_2} } + 
\frac{\AA{\ub_1 u_1} \, \AA{\ub_1 \ub_2} \, \SSS{\ub_1 u_2} }{\kstr_{u_1} \, \AL{\ub_1}\slashk_{u_1} \SR{\ub_1} \, \AA{\ub_2 u_2}\, \SSS{\ub_2 u_2}\, }
\nn \\ 
&+&
\frac{\AA{u_1 \ub_2} \, \SSS{ \ub_1 u_1} \, \SSS{u_1 u_2} }{\kapp_{\ub_1} \, \AL{u_1}\slashk_{\ub_1} \SR{u_1} \, \AA{\ub_2u_2}\, \SSS{\ub_2 u_2}\, } \, , 
\nn \\
\Amp( \bar{u_1}^{*+},u_2^-,u_1^{*-},\ub_2^+) 
&=& -
\frac{ \SSS{\ub_1 \ub_2} \, \AA{u_1 u_2} }{ \kapp_{\ub_1}\, \kstr_{u_1}\, \left( k_{u_1} + p_{\ub_2} \right)^2} +
\frac{ \SSS{\ub_1 \ub_2}\, \AA{\ub_1 u_1}}{\kstr_{u_1} \, \left( k_{u_1}+p_{\ub_2} \right)^2  \SSS{\ub_1 u_2}}
\nn \\
&-&
\frac{ \AA{ u_1 u_2 }\, \SSS{\ub_1 u_1}}{\kapp_{\ub_1} \, \left( k_{u_1}+p_{\ub_2} \right)^2  \AA{u_1 \ub_2 }} -
\frac{ \AA{\ub_1 u_1} \, \SSS{\ub_1 u_1} }{ \left( k_{u_1} + p_{\ub_2} \right)^2 \, \AA{u_1\ub_2}\, \SSS{ \ub_1 u_2} } \, , 
\nn \\
\Amp( \bar{u}_1^{*+},u_2^-,u_1^{*+},\ub_2^-) 
&=& -
\frac{ \SSS{\ub_1 u_1} \, \AA{\ub_2 u_2} }{ \kapp_{\ub_1}\, \kapp_{u_1}\, \left( k_{u_1} + p_{\ub_2} \right)^2} +
\frac{ \SSS{\ub_1 u_1}\, \AA{u_1 u_2}}{\kapp_{\ub_1} \, \left( k_{u_1} + p_{\ub_2} \right)^2  \SSS{u_1\ub_2}} 
\nn \\
&-&
\frac{ \AA{ \ub_1 \ub_2  }\, \SSS{\ub_1 u_1}}{\kapp_{u_1} \, \left( k_{u_1}+p_{\ub_2} \right)^2  \SSS{\ub_1 u_2} } +
\frac{ \AA{\ub_1 u_1} \, \SSS{\ub_1 u_1}}{ \left( k_{u_1} + p_{\ub_2} \right)^2 \, \SSS{u_1\ub_2}\, \SSS{\ub_1u_2} } \, .
\eea
Here the labels $1$ and $2$ stand respectively for the off-shell and on-shell quark-antiquark pair, whereas the t-channel is made
by the momenta of the two (anti-) quarks.

The interested reader can reproduce these results using the Feynman rules 
and applying them to Feynman diagrams which are similar to the ones shown in the previous section. 
Specifically, the amplitudes $\Amp(u_1^{*+},\ub_1^-,\ub_2^-,u_2^{*+}) $
$\Amp(u_1^{*+},\ub_1^-,\ub_2^+,u_2^{*-})$ are $s$-channel processes and are thus represented by diagrams
similar to the set in Fig. \ref{Fig_offubuondbd}. The four amplitudes $\Amp(u_1^{*+},\ub_2^-,\ub_1^-,u_2^{*+})$, 
$\Amp(u_1^{*+},\ub_2^-,\ub_1^+,u_2^{*-})$, $\Amp( \bar{u}_1^{*+},u_1^{*-},\ub_2^+,u_2^-) $ and $\Amp( \bar{u}^{*+}_1,u_1^{*-},\ub_2^-,u_2^+)$
are $t$- channel processes; $\Amp( \bar{u_1}^{*+},u_2^-,u_1^{*-},\ub_2^+)$ and $\Amp( \bar{u}_1^{*+},u_2^-,u_1^{*+},\ub_2^-)$
are $u$-channel scatterings. Each of these later six processes is made of four contributions analogous to those 
depicted in Fig. \ref{Fig_offubdbonud}, with the only different that the gluon exchange must happen in the corresponding 
$t$ or $u$ channel instead than in the $s$ one.

\section{The general formulas for MHV amplitudes}\label{MHV}

In this section we derive formulas for amplitudes which are Maximally Helicity Violating (MHV).
First of all we recall the MHV formulas derived in \cite{vanHameren:2014iua} for the all-gluon case,
\bea
\Amp(1^*,i^*,\left( \text{the rest}^+ \right))
&=&
\frac{1}{\kstr_1 \kstr_i} \, \frac{\AA{1 i}^4}{\AA{12}\AA{23}\dots\AA{n-2 | n-1}\AA{n-1|n}\AA{n1}}
\nn \\
\Amp(1^*,i^*,\left( \text{the rest}^- \right))
&=&
\frac{1}{\kapp_1 \kapp_i} \, \frac{\SSS{i 1}^4}{\SSS{1n}\SSS{n|n-1}\SSS{n-1 | n-2}\dots\SSS{32}\SSS{21}} \, .
\label{MHV_gluons}
\eea

Then we set up to find out the general structure of the MHV amplitudes in the case we have
1 off-shell gluon and 1 off-shell fermion and two off-shell fermions.

For the first case, the possible MHV amplitudes are
$\Amp(1^*,2^+,\dots,n^+,\qb^*,q^+)$ and $\Amp(1^*,2^+,\dots,n^+,\qb^+,q^*)$
plus their conjugated amplitudes.

Choosing $e^\mu = \AL{\qb} \gamma^\mu \SR{1}/2$ and $e^\mu = \AL{q} \gamma^\mu \SR{1}/2$, 
we find that all residues but $D^1$ vanish identically, leading straight to
\bea
\Amp(1^*,2^+,\dots,n^+,\qb^*,q^+) 
&=&
\frac{1}{x_1\,\kstr_1} \, \Amp(\hat{1}^-,2^+,\dots,n^+,\hat{\qb}^*,q^+) =
\frac{1}{\kstr_1\, \kstr_{\qb}} \, \frac{ \AA{1\qb}^3 \, \AA{1q} }{\AA{12} \, \dots \, \AA{\qb q} \, \AA{q1} } \, ,
\nn \\
\Amp(1^*,2^+,\dots,n^+,\qb^+,q^*) 
&=&
\frac{1}{x_1\,\kstr_1} \, \Amp(\hat{1}^-,2^+,\dots,n^+,\qb^+,\hat{q}^*) =
\frac{1}{\kstr_1\,\kstr_{q}} \, \frac{ \AA{1q}^3 \, \AA{1\qb} }{\AA{12} \, \dots \, \AA{\qb q} \, \AA{q1} } \, ,
\nn \\
\eea
where the last steps are taken through Table \ref{CDg}.
It is obvious that their conjugated amplitudes are just
\bea
\Amp(1^*,2^-,\dots,n^-,\qb^*,q^-) 
&=&
\frac{1}{\kapp_1\, \kapp_{\qb}} \, \frac{ \SSS{1\qb}^3 \, \SSS{1q} }{ \SSS{q\qb} \, \dots \, \SSS{21} \, \SSS{1q} } \, ,
\nn \\
\Amp(1^*,2^-,\dots,n^-,\qb^-,q^*) 
&=&
\frac{1}{\kapp_1\,\kapp_{q}} \, \frac{ \SSS{1q}^3 \, \SSS{1\qb} }{\SSS{q\qb} \, \dots \, \SSS{21} \, \SSS{1q} } \, .
\eea

Then the case of the off-shell fermion pair. We start with $\Amp(1^+,2^+,\dots,n^+,\qb^{*+},q^{*-})$.
Because of the results derived in the previous section, we assume the inductive hypothesis 
\beq
\Amp(1^+,2^+,\dots,n^+,\qb^{*+},q^{*-}) = \frac{1}{\kstr_q} \, \frac{\AA{q\qb}^3}{\AA{12} \dots \AA{\qb q} \, \AA{q1} }
\label{inductive}
\eeq
and we choose $e^\mu = \AL{q}\gamma^\mu\SR{1}/2$, so that the $C^q$ residue vanishes and we have
\bea
\Amp(1^+,2^+,\dots,n^+,\qb^{*+},q^{*-})
&=&
\Amp(\hat{1}^+,2^+,\hat{K}^-)\, \frac{1}{\AA{12}\,\SSS{21}} \, \Amp(-\hat{K},3^+,\dots,\qb^{*+},q^{*-}) 
\nn \\
&&\hspace{-50mm}
=
\frac{\SSS{21}^3}{\SSS{1\hat{K}} \SSS{\hat{K}2} }\, \frac{1}{\AA{12}\SSS{21}} \, \frac{1}{\kstr_q} \, \frac{\AA{q\qb}^3}{ \AA{\hat{K}3} \, \AA{34} \dots \AA{\qb q}\,\AA{q\hat{k}} }
= \frac{1}{\kstr_q} \, \frac{\AA{q\qb}^3 \, \SSS{12}^2}{ \AA{12}\dots \AA{\qb q} \, \AL{q} \hat{\slashK} \SR{2} \, \AL{3} \hat{\slashK} \SR{1}}
\nn \\
&=&
\frac{1}{\kstr_q} \, \frac{\AA{q\qb}^3}{\AA{12} \dots \AA{\qb q} \, \AA{q1} } \, ,
\label{MHV_2fer_1}
\eea
where a possible $B$ residue drops off because of the all-plus sub-amplitude on the left, which is of course null. 
The $A$ residue is worked out using the inductive hypothesis (\ref{inductive}).
Similarly, using $e^\mu = \AL{\qb}\gamma^\mu\SR{1}/2$, in the same way one gets
\bea
\Amp(1^+,2^+,\dots,n^+,\qb^{*-},q^{*+}) 
&=&  
\frac{1}{\kstr_{\qb}} \, \frac{\AA{q\qb}^3}{\AA{12} \dots \AA{\qb q} \, \AA{q1} } \, ,
\label{MHV_2fer_2}
\eea
%
Of course the adjoints of (\ref{MHV_2fer_1}) and (\ref{MHV_2fer_2}) are
\bea
\Amp(1^-,2^-,\dots,n^-,\qb^{*-},q^{*+}) 
&=& 
\frac{1}{\kapp_q} \, \frac{\SSS{\qb q}^3}{ \SSS{q\qb} \dots \SSS{21} \, \SSS{1q} } \, ,
\nn \\
\Amp(1^-,2^-,\dots,n^-,\qb^{*+},q^{*-}) 
&=& 
\frac{1}{\kapp_{\qb}} \, \frac{\SSS{\qb q}^3}{ \SSS{q\qb} \dots \SSS{21} \, \SSS{1q} } \, .
\eea
%

\section{Squared $2 \rightarrow 2$ matrix elements}\label{Squared}%

In this section we sketch the kinematical properties which are specific
to matrix elements in HEF and may be relevant for further phenomenological investigations.

We also square our matrix elements by averaging (summing) over the initial (final) state colours and spins,
explaining how to express them  in terms of the rapidity difference $\Delta y$ and the azimuthal angle difference 
$\Delta\phi$ between the final state partons. Only for this section, we adopt the (physical) convention that the 
two off-shell particles are incoming and the other two are outgoing. 
This means that momentum conservation reads
\beq
k_a + k_b = p_a + p_b \, .
\eeq
In doing this, we generalise the discussion 
of the collinear case presented in~\cite{DelDuca:1995hf}

\begin{figure}[h]
\begin{center}
\includegraphics[scale=0.45]{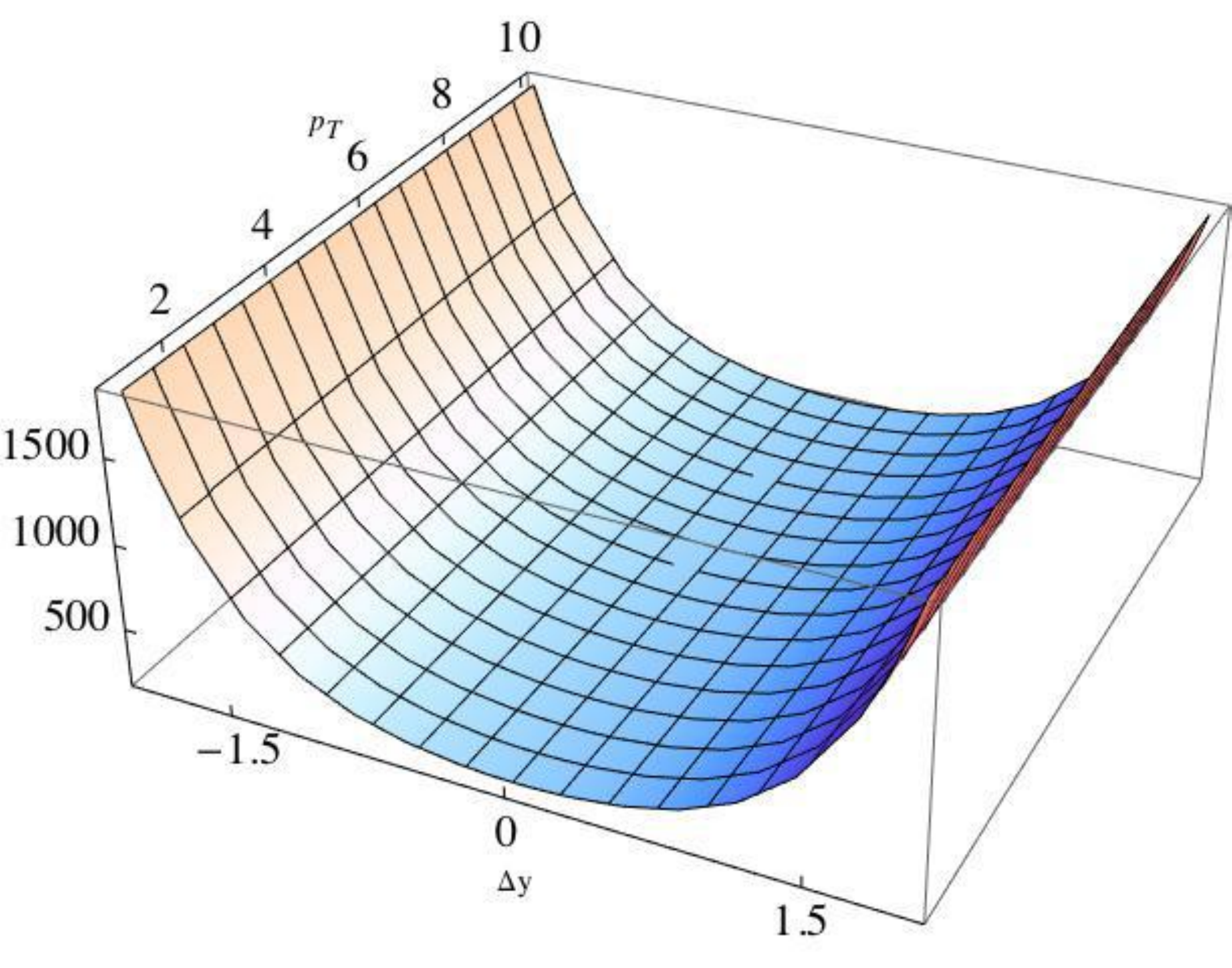}
\caption{$\left|M_{g_1 g_2 \rightarrow g_3 g_4} \right|^2$}
\label{Fig_squaredplot_1}
\end{center}
\end{figure}

\begin{figure}[h]
\begin{center}
\includegraphics[scale=0.60]{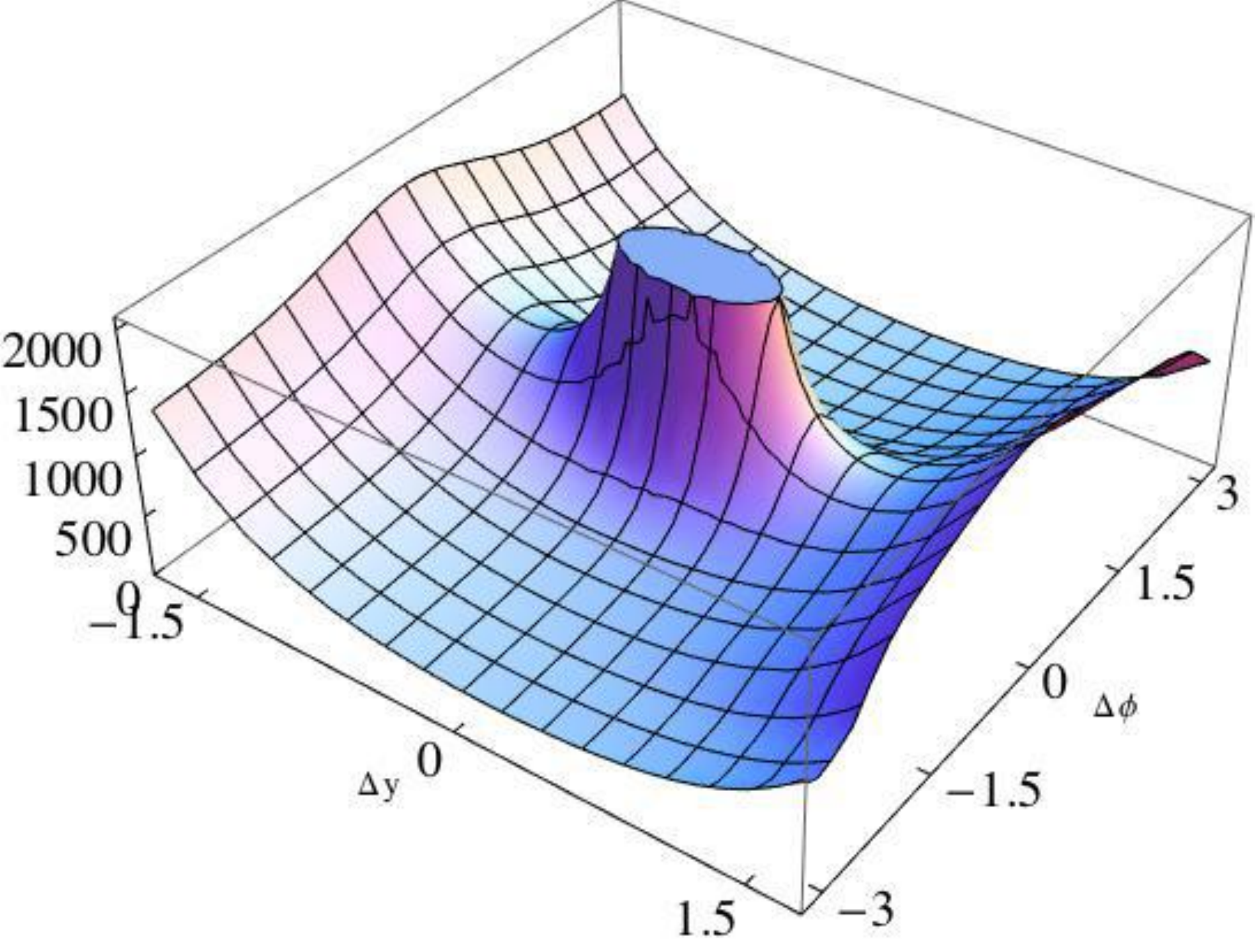}
\caption{$\left|M_{g^*_1 g_2 \rightarrow g_3 g_4} \right|^2$ for $k_{3\perp} = k_{4\perp} = 20$ GeV}
\label{Fig_squaredplot_2}
\end{center}
\end{figure}

\begin{figure}[h]
\begin{center}
\includegraphics[scale=0.60]{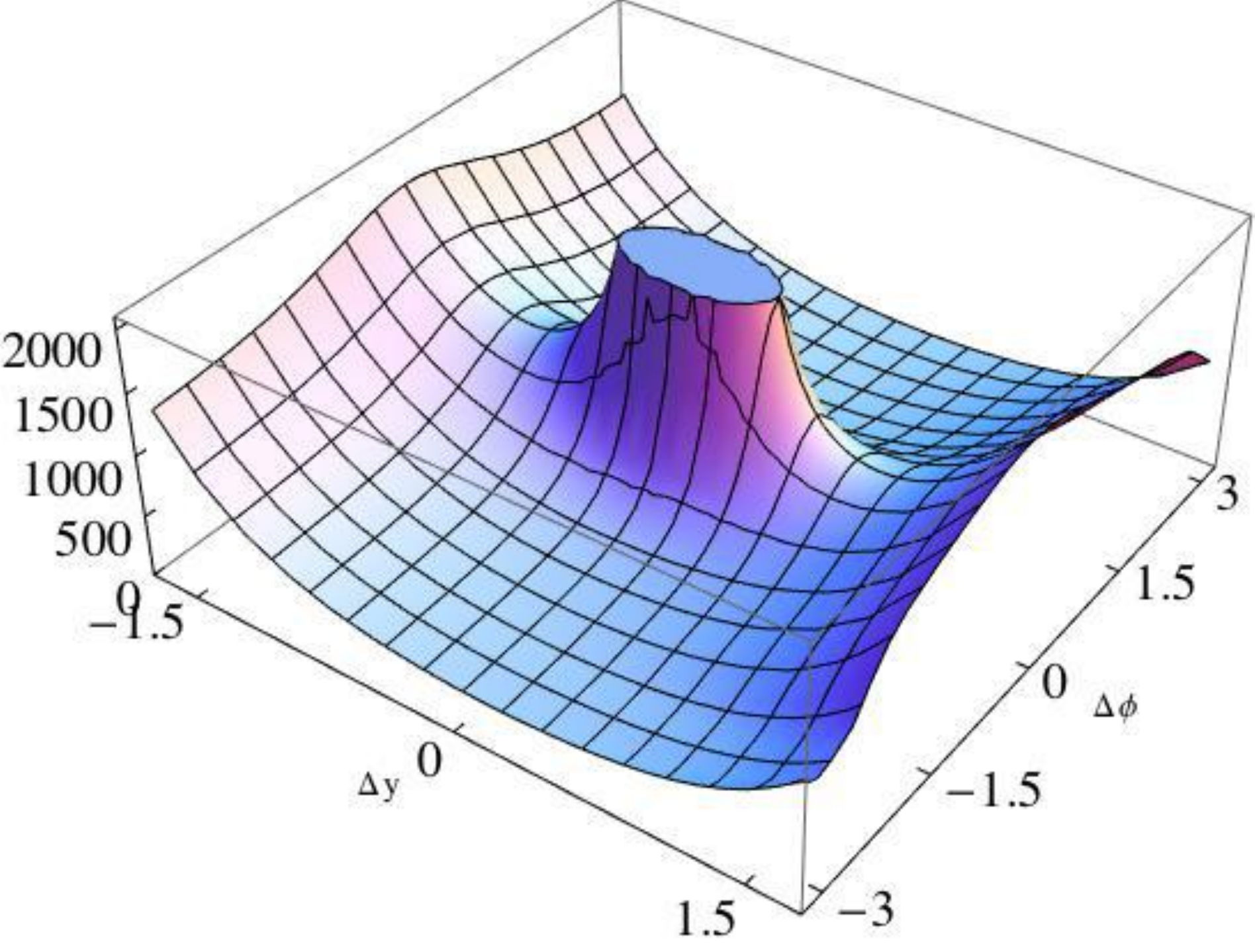}
\caption{$\left|M_{g^*_1 g^*_2 \rightarrow g_3 g_4} \right|^2$ for
$p_{1\perp} = k_{3\perp} = k_{4\perp} = 20 $ GeV,  $\theta_1 = 0.2, y_1 = - 2, \phi_1 = 2.9$}
\label{Fig_squaredplot_3}
\end{center}
\end{figure}

\begin{figure}[h]
\begin{center}
\includegraphics[scale=0.60]{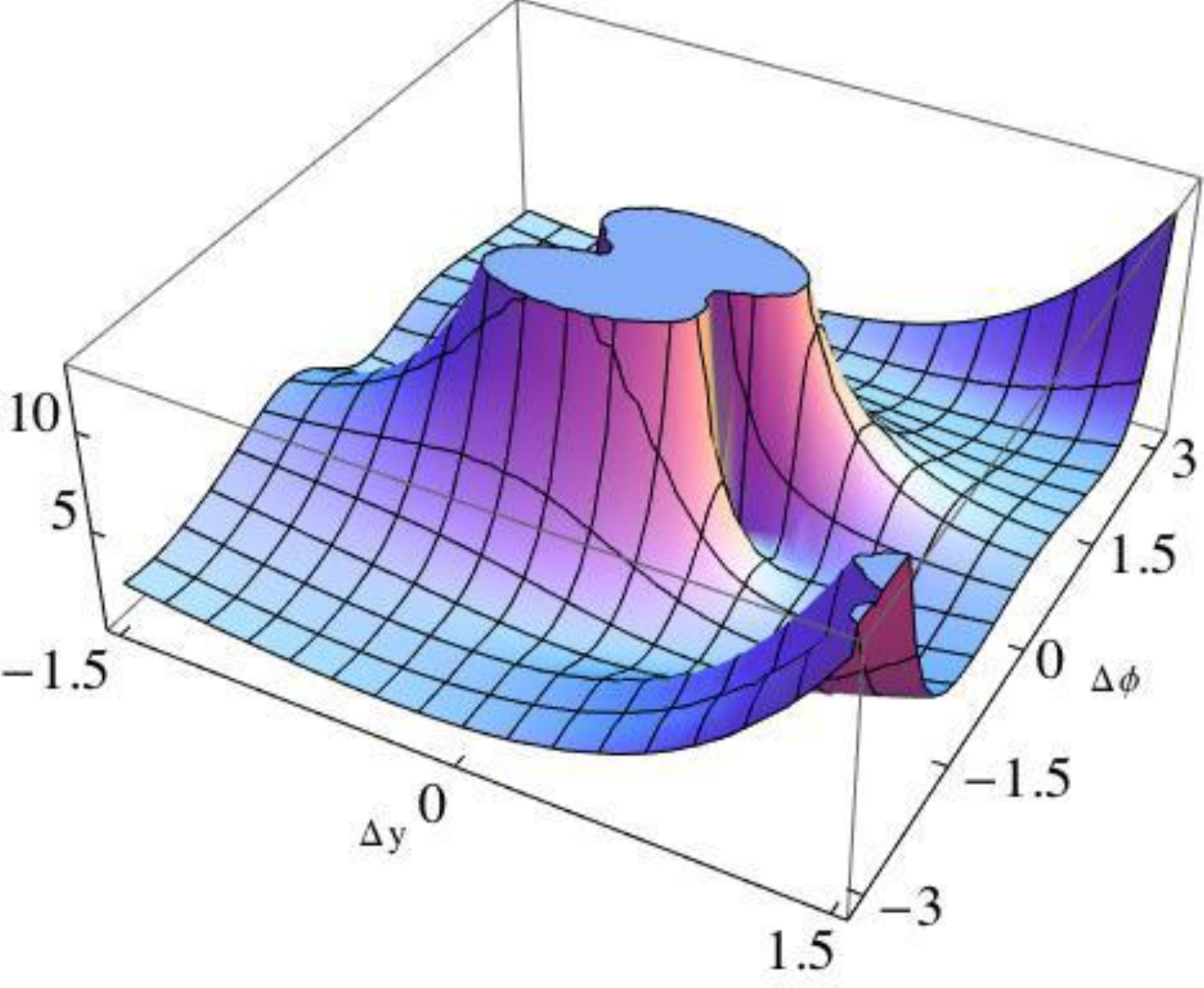}
\caption{$\left|M_{g^*_1 g^*_2 \rightarrow q \qb}\right|^2$ for 
$p_{1\perp} = k_{q\perp} = k_{\qb\perp} = 20 $ GeV,  $\theta_1 = 0.2, y_1 = - 2, \phi_1 = 2.9$}
\label{Fig_squaredplot_4}
\end{center}
\end{figure}

Working in light-cone variables, the $q^\pm$ momenta for a general 4-vector $q$ are defined as
\beq
q^{\pm} = q^0 \pm q^1 \, ,
\label{momcon2to2}
\eeq
and $x_1$ is the beam direction.
The rapidity variable is defined as
\beq
y = \frac{1}{2}\, \log \left(\frac{q^+}{q^-}\right) \Leftrightarrow \tanh y = \frac{q^1}{q^0} \, ,
\eeq
which naturally allows to express the particle energy and longitudinal momentum of massless particles as
\bea
q^0
&=& q_\perp \, \sinh y
\nn \\
q^1
&=& q_\perp \, \cosh y
\eea
The incoming momenta $k_{a/b}$ and
the outgoing momenta $p_{a/b}$ are thus found to be respectively
\bea
k_a &=& \left(  \sqrt{S}\, x_a, 0 ; \vec{k}_{a\perp}  \right) \nn \\
k_b &=& \left(  0, \sqrt{S}\, x_b ; \vec{k}_{b\perp}  \right) \nn \\
p_{a} &=& \left( p_{a\perp} e^{y_a}, p_{a\perp} e^{-y_a}; \vec{p}_{a\perp} \right) = p_{a\perp} \, \left(  e^{y_a}, e^{-y_a}; \cos \phi_a, \sin\phi_a \right) \nn \\
p_{b} &=& \left( p_{b\perp} e^{y_b}, p_{b\perp} e^{-y_b}; \vec{p}_{b\perp} \right) = p_{b\perp} \, \left(  e^{y_b}, e^{-y_b}; \cos \phi_b, \sin\phi_b \right) \, .
\label{param1}
\eea
Here, differently from the collinear case, $\vec{k}_{a/b\perp} \neq 0$, which implies imbalance for the final state transverse momenta, 
$\vec{p}_{a\perp} \neq - \vec{p}_{b\perp} $; $k_{a/b\perp} \equiv |\vec{k}_{a/b\perp}|$, $p_{a/b\perp} \equiv |\vec{p}_{a/b\perp}|$,
$S$ is the proton center of mass energy, $y_{a/b}$ are the rapidities of the final states particles and 
the azimuthal angle difference is defined as the angle between their directions 
$\Delta\phi = \arccos\left(\hat{p}_{a\perp}\, \cdot \hat{p}_{b\perp}\right) $, 
whereas $\theta_a$ is the azimuthal angle for the transverse momentum of the first incoming off-shell particle.
By projecting momentum conservation (\ref{momcon2to2}) onto the light-cone components, one easily gets the further constraints
\bea
x_{a} &=& \frac{p_{a\perp} e^{y_a} + p_{b\perp} e^{y_b}}{\sqrt{S}} 
\nn \\
x_{b} &=& \frac{p_{a\perp} e^{-y_a} + p_{b\perp} e^{-y_b}}{\sqrt{S}} \, .
\label{xconst}
\eea
By using (\ref{xconst}), we finally end up with the parameterizations
\bea
k_a &=& \left(  p_{a\perp} e^{y_a} + p_{b\perp} e^{y_a+\Delta y}, 0 ; k_{a\perp} \cos \theta_a, k_{a\perp} \sin \theta_a  \right) \, , \nn \\
%
%
p_{a} &=&  p_{a\perp} \, \left(  e^{y_a}, e^{-y_a}; \cos \phi_a, \sin\phi_a \right) \, ,\nn \\
p_{b} &=&  p_{b\perp} \, \left(  e^{y_a + \Delta y}, e^{-y_a - \Delta y}; \cos(\phi_a + \Delta\phi), \sin(\phi_a + \Delta\phi) \right)\, , \nn \\
k_b &=& -k_a + p_a + p_b \, ,
\label{param2}
\eea
after introducing the rapidity difference $\Delta y$.
We are thus left with $8$ independent parameters on which the squared matrix elements depend:
$ k_{a\perp}, \theta_a, p_{a\perp}, p_{b\perp}, y_a, \phi_a, \Delta y, \Delta \phi $.
At the end of this paper, we present some plots of squared matrix elements as functions of $(\Delta y,\Delta \Phi)$
for the following given values of the first $6$ variables,
$k_{a\perp} = p_{a\perp} = p_{b\perp} = 20 $ GeV,  $\theta_a = 0.2, y_a=-2, \phi_a=2.9$.
We choose to show the processes $g^*\, g^* \rightarrow g g $, $g^*\, g^* \rightarrow q \qb $ and $q\, q \rightarrow q q $ 
plotted in Figs. \ref{Fig_squaredplot_3}-\ref{Fig_squaredplot_4}.
In Figs. \ref{Fig_squaredplot_1}-\ref{Fig_squaredplot_2}, instead, we have inserted the plots 
of the squared amplitudes for 2-gluon scattering with two and one on-shell legs respectively.
As momentum conservation in the collinear case enforces $\phi = \pi$, the second variable is chosen
to be the final state transverse momentum of each jet, so that the plot of an actual process is to be thought
as a one dimensional section perpendicular to the $k_T$ axis. It is nevertheless useful for illustrative purposes.


\begin{figure}[h]
\begin{center}
\includegraphics[scale=0.50]{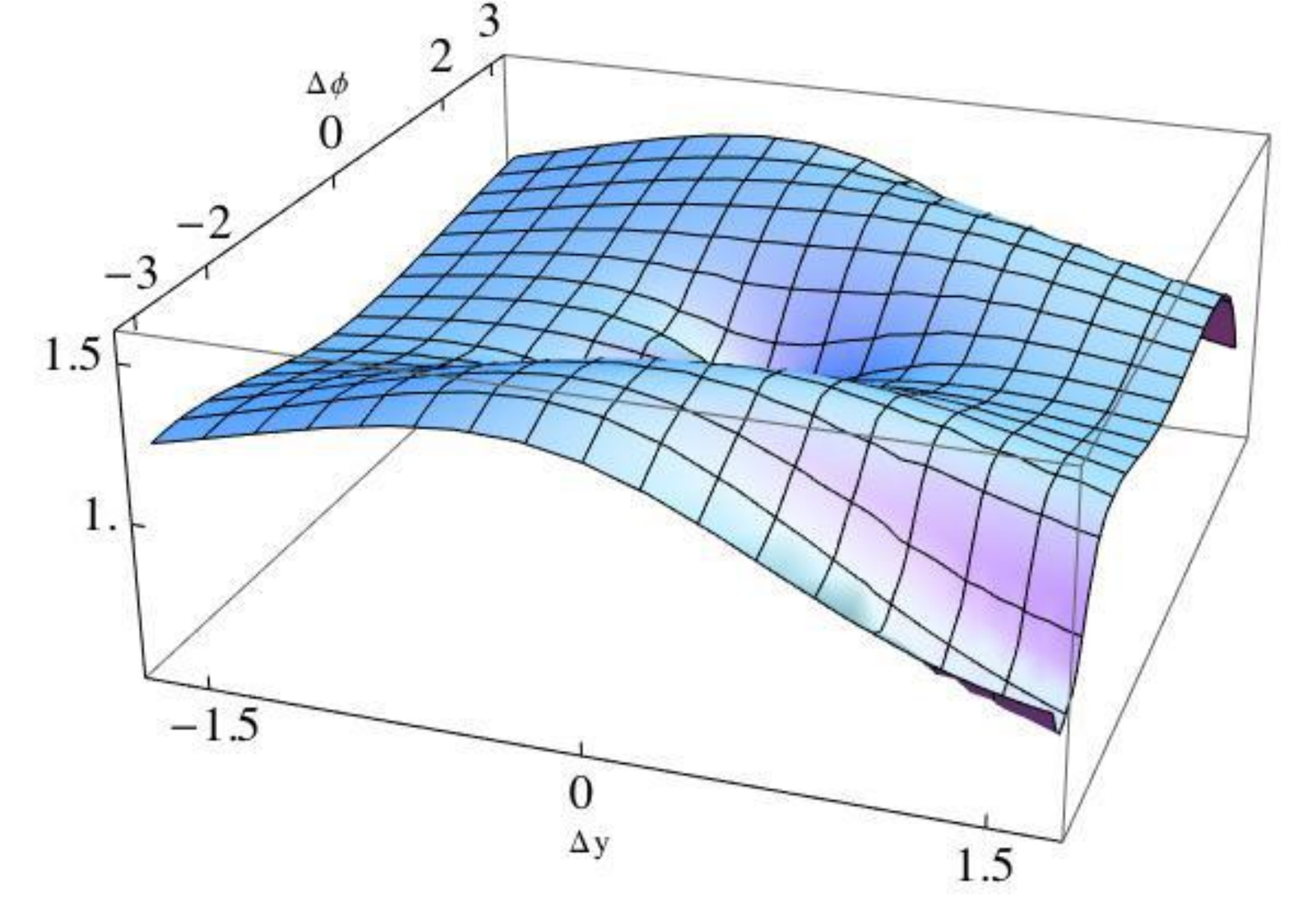}
\caption{$\left|\frac{M_{g^*_1 g^*_2 \rightarrow g_3 g_4}}{M_{g^*_1 g_2 \rightarrow g_3 g_4}}\right|^2$ for 
$\theta_1 = 0.2, y_1 = - 2, \phi_1 = 2.9$ and $p_{1\perp} = k_{3\perp} = k_{4\perp} = 20 $ GeV,   for the matrix element with two off-shell legs.}
\label{Fig_4g_comparison_1}
\end{center}
\end{figure}

\begin{figure}[h]
\begin{center}
\vspace{20mm}
\includegraphics[scale=0.50]{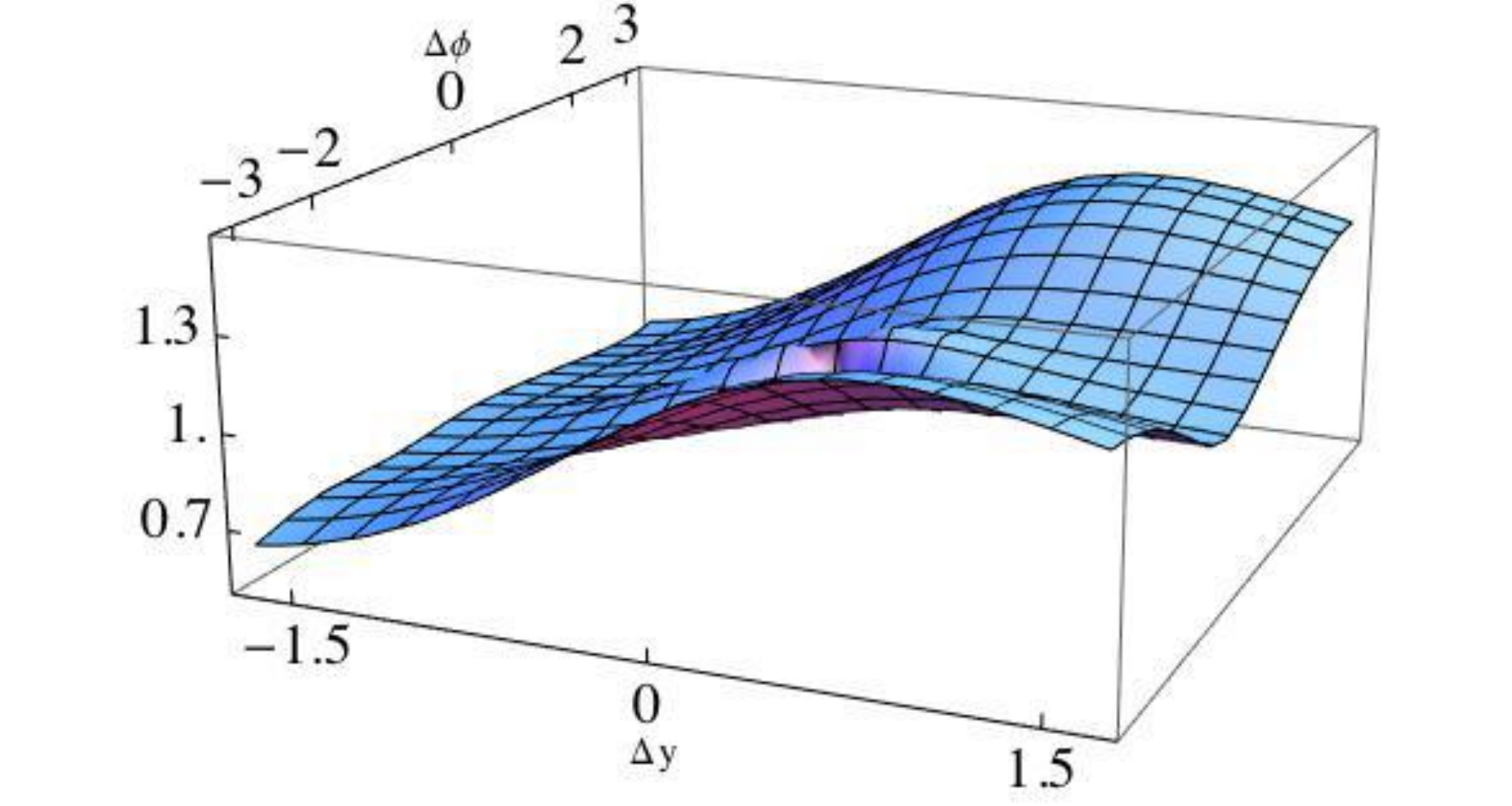}
\caption{$\left|\frac{M_{g^*_1 g^*_2 \rightarrow g_3 g_4}}{M_{g^*_1 g_2 \rightarrow g_3 g_4}}\right|^2$ for
$\theta_1 = 0.2, y_1 = - 2, \phi_1 = 2.9$ and $p_{1\perp} = 20$ GeV, $k_{3\perp} = 30$ GeV , $k_{4\perp} = 40 $ GeV for the matrix element with two off-shell legs.}
\label{Fig_4g_comparison_2}
\end{center}
\end{figure}

Finally, in Figs. \ref{Fig_4g_comparison_1} and \ref{Fig_4g_comparison_2} we show explicitly the ratio of the squared matrix elements for
$g^*\, g^* \rightarrow g g $ and $g^*\, g \rightarrow g g $ for the aforementioned as well as one more 
set of input values, which differ from the previous ones only for the value of $\phi_a = 0.5$; 
this is done to stress the non trivial difference between the two configurations and the sensitivity of the difference between the two cases to initial conditions.
In all the plots the strong coupling constant is set $ \alpha_s = 0.2$, so as to facilitate the comparison with the plots in \cite{Kotko:2015ura}.
The partial amplitudes making up the matrix element for $g^* g \rightarrow g g$, which are all MHV, can be found in \cite{vanHameren:2014iua,vanHameren:2015bba}. 

For the sake of clarity, we will limit our discussion to the only-gluon case. 
What is striking in Figs. \ref{Fig_squaredplot_2} and \ref{Fig_squaredplot_3} 
with respect to Fig. \ref{Fig_squaredplot_1} is the possibility of having the configuration of final state jets away from back-to-back configuration.
These configurations are precisely what High Energy Factorization is supposed to produce, thanks to the transverse
momentum imbalance due to the initial state partons non vanishing $k_T$. 


\begin{figure}[h]
\begin{center}
\vspace{20mm}
\includegraphics[scale=0.50]{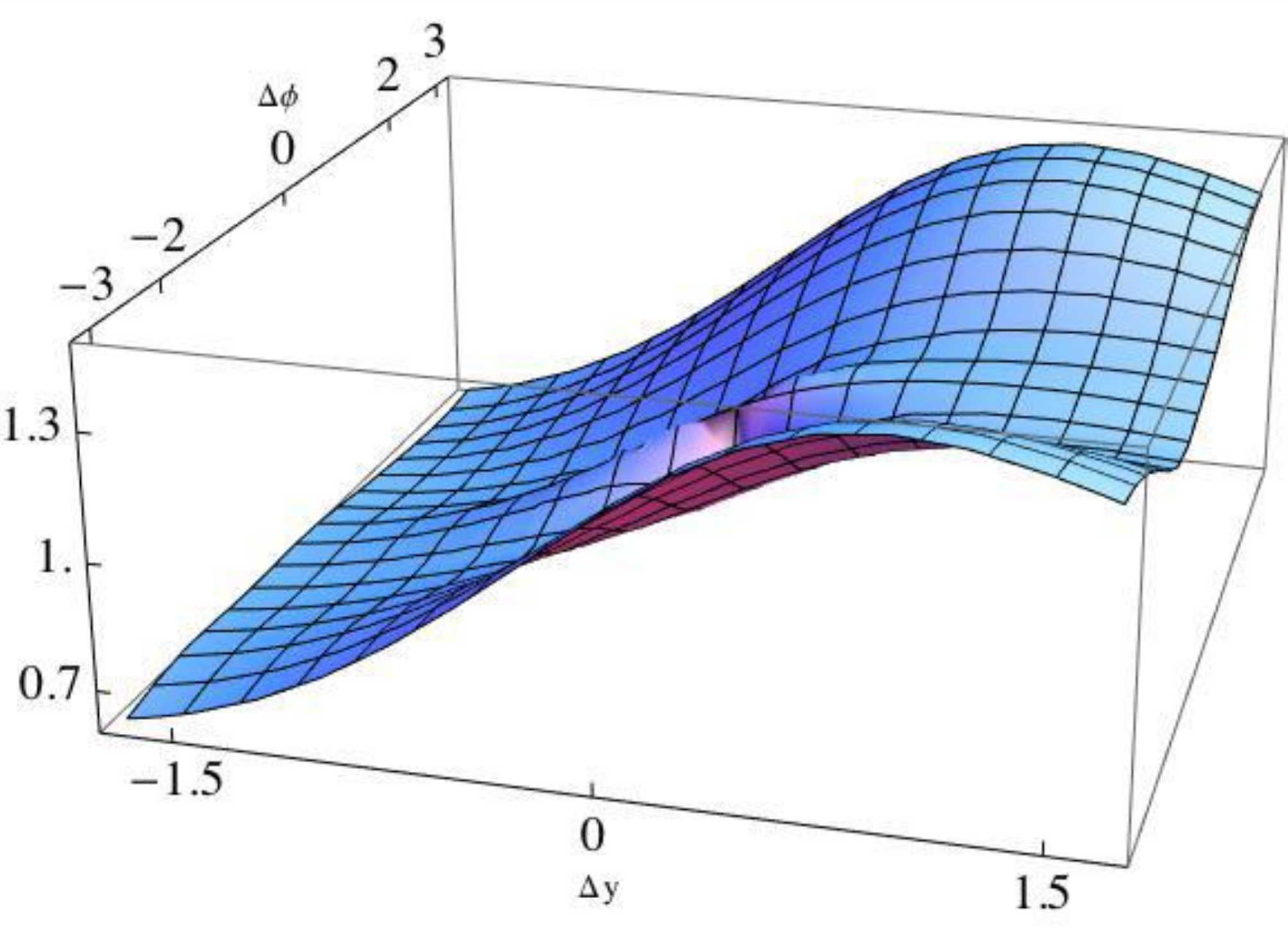}
\caption{$\left|\frac{M_{g^*_1 g^*_2 \rightarrow g_3 g_4}}{M_{g^*_1 g_2 \rightarrow g_3 g_4}}\right|^2$ for
$\theta_1 = 0.2, y_1 = - 2, \phi_1 = 2.9$ and $p_{1\perp} = 20$ GeV, $k_{3\perp} = 30$ GeV , $k_{4\perp} = 50 $ GeV for the matrix element with two off-shell legs.}
\label{Fig_4g_comparison_3}
\end{center}
\end{figure}

In the collinear approach, such a dispersion for $2\rightarrow 2$ LO scattering process 
is generated at the cross section level by initial state parton shower effects.
The spike in the center, which diverges for vanishing $\Delta\phi$ and $\Delta y$, 
corresponds to the physical situation in which the two jets originating from the final state partons are unresolved and become one single jet. \\
As for the comparison between the hybrid factorization case (only one off-shell leg) and the full High Energy Factorization one (two off-shell legs),
we limit ourselves to showing three plots - Figs. \ref{Fig_4g_comparison_1}-\ref{Fig_4g_comparison_3} - 
featuring the ratios of only-gluon squared matrix elements on a coinciding subset of independent variables 
(i.e. the final state transverse momenta), while keeping the other variables ($\theta_1$, $\phi_1$, $y_1$ and $p_{T1}$) fixed. 
As we scan three possible differences of final state transverse momenta, in ascending order,
we clearly observe that the rapidity gradient in the ratio plots increases.
A more complete scan of the various possible kinematic configurations would require too much space for a
paper like the present one, which is concerned mainly with theoretical aspects of the problem,
but remains an interesting problem to be addressed separately.

%

\section{Summary and perspectives}\label{summary}%

We have studied the problem of calculating amplitudes with 2 off-shell legs for High Energy Factorization via BCFW recursion.
We found a full generalisation of the recursibility criteria devised in \cite{vanHameren:2014iua,vanHameren:2015bba},
proving independently that, exactly as in the fully on-shell case, amplitudes with 
at least one gluon leg can be recursed, whereas, when there are only external fermions, 
one still has to rely on the ordinary Feynman diagram expansion. Other types of generalisations 
of BCFW are underway in the on-shell case~\cite{Jin:2015pua,Feng:2014pia,Jin:2014qya,Feng:2011twa},
which amount to devise ways to pin down the boundary contributions which show up when the asymptotic
condition on the complexified amplitude is not satisfied. The off-shell implementation of such methods is,
however, well beyond the scope of the present work and we will not discuss it further.

For the first time, a complete set of exact analytical expressions of color ordered 4-point amplitudes
with 2 off-shell legs for a Yang-Mills theory with massless fermions has been provided,  
together with MHV formulas for any number of legs. 
Also a discussion of their properties and differences w.r.t. the collinear case is proposed.
All expressions for the amplitudes, except the cases with two quark-antiquark pairs, have been implemented in {\sc amp4hef}~\cite{Bury:2015dla} for numerical evaluation.

Concerning the potential future developments, it would of course be good to push the program to higher multiplicities, especially
now that the theoretical framework to perform such calculations has been completely defined. Considering the 
present applicability limits mentioned at the end of section \ref{res}, it would be nice to arrive at expressions for 
$6$- or $7$-point amplitudes. Nevertheless, performing such calculations by hand for every case 
can be quite demanding, so the development of a dedicates software package is envisaged to this goal. \\

{\bf Acknowledgments}

We thank M.~A.~Nefedov, A.~V.~Shipilova and V.~A.~Saleev for useful correspondence.
MS and KK are supported by the NCN grant DEC-2013/10/E/ST2/00656.
AvH is supported by the grant of the National Science Center, Poland, No. 2015/17/B/ST2/01838.
All figures were drawn with \emph{feynMF}~\cite{Ohl:1995kr}.


\appendix

\section{The 3-point amplitudes with less off-shell legs}\label{3less}

In order to make our paper self-consistent, here we list 
the 3-point amplitudes used in the construction of 
the 4-point amplitudes presented in this paper.
They are either on-shell or have only 1 off-shell leg.

\subsection{On-shell 3-point amplitudes}

%
\begin{eqnarray}
\Amp(1^+,2^-,3^-) = \frac{\AA{23}^3}{\AA{12}\AA{31}} \, 
\quad
&&
\Amp(1^-,2^+,3^+) = \frac{\SSS{32}^3}{\SSS{21}\SSS{13}} \nn
\\
\Amp(1^-,\qb^+,q^-) = \frac{\AA{1 q}^3\AA{1 \qb}}{\AA{1 \qb}\AA{\qb q}\AA{q 1}}\, 
\quad
&&
\Amp(1^+,\qb^+,q^-) = \frac{\SSS{1 \qb}^3\SSS{1 q}}{\SSS{1 q}\SSS{q \qb}\SSS{\qb 1}} \nn
\\
\Amp(1^-,\qb^-,q^+) = \frac{\AA{1 \qb}^3\AA{1 q}}{\AA{1 \qb}\AA{\qb q}\AA{q 1}}
\quad
&&
\Amp(1^+,\qb^-,q^+) = \frac{\SSS{1 q}^3\SSS{1 \qb}}{\SSS{1 q}\SSS{q\qb}\SSS{\qb 1}} \nn
\end{eqnarray}
%

\subsection{3-point amplitudes with 1 off-shell leg}

%
\bea
&&
\Amp(1^*,2^+,3^-) = \frac{1}{\kstr_1}\frac{\AA{31}^3}{\AA{12}\AA{23}} = \frac{1}{\kappa_1}\frac{\SSS{21}^3}{\SSS{13}\SSS{32}}   \, ,
\\
&&
\Amp(1^\pm,2^\pm,3^*) =  0  \, ,
\nn \\
&&
\Amp(1^*,\qb^+,q^-)  = \frac{1}{\kappa_1}\, \frac{\SSS{1 \qb}^3 \SSS{1 q} }{\SSS{1 q} \SSS{q \bar{q} } \SSS{\bar{q}  1} }  = 
\frac{1}{\kappa^*_1}\, \frac{\AA{1 q}^3 \AA{1 \qb} }{\AA{1 \bar{q}  } \AA{ \bar{q} q } \AA{q 1}   }
\\
&&
\Amp(1^*,\qb^-,q^+)  = 
\frac{1}{\kappa_1}\, \frac{\SSS{1 q}^3 \SSS{1 \qb} }{\SSS{1 q} \SSS{q \bar{q} } \SSS{\bar{q}  1} }  = 
\frac{1}{\kappa^*_1}\, \frac{\AA{1 \qb}^3 \AA{1 q} }{\AA{1 \bar{q}  } \AA{ \bar{q} q } \AA{q 1}   } 
\\
&&
\Amp(1^+,\qb^*,q^-) =  \frac{1}{\kappa_{\qb}}\, \frac{ \SSS{1 \qb}^3 \SSS{1 q} }{ \SSS{1 q} \SSS{q \qb} \SSS{\qb 1} }  \, ,
\quad
\Amp(1^-,\qb^*,q^+) =  \frac{1}{\kstr_{\qb}}\, \frac{ \AA{1 \qb}^3 \AA{1 q} }{\AA{1 \qb} \AA{\qb q} \AA{q 1} } \, .
\\
&&
\Amp(1^+,\qb^-,q^*) =  \frac{1}{\kappa_{q}}\, \frac{ \SSS{1 q}^3 \SSS{1 \qb} }{ \SSS{1 q} \SSS{q \qb} \SSS{\qb 1} }  \, ,
\quad
\Amp(1^-,\qb^+,q^*) =  \frac{1}{\kstr_{q}}\, \frac{ \AA{1 q}^3 \AA{1 \qb} }{\AA{1 \qb} \AA{\qb q} \AA{q 1} } \, ,
\\
&&
\Amp(1^+,\qb^*,q^+)  =  0
\quad
\Amp(1^-,\qb^*,q^-)  = 0
\label{exception_1}
\\
&&
\Amp(1^+,\qb^+,q^*)  =  0
\quad
\Amp(1^-,\qb^-,q^*)  =  0
\label{exception_2}
\eea
%

\section{The color-order reversed relation for gluon amplitudes}\label{Feng}

A fundamental property of gluon on-shell 3-point amplitudes is that $\Amp(1,2,3) = - \Amp(3,2,1)$,
which is preserved also for the amplitudes with 1 and 2 off-shell legs.
In \cite{Feng:2010my} this allows to propose the inductive hypothesis
$\Amp(1,2,\dots,n-1, n) = (-1)^n\, \Amp(n,n-1,\dots,2,1)$.

Then one uses BCFW with shift vector $\AL{1}\gamma^\mu\SR{n}/2$ to express 
the general color-reversed n-point amplitude as
\bea
\Amp(n,n-1,\dots,2,1) 
&=&
\sum_{i=1,n-3} \Amp(\hat{n},n-1,\dots,i,-\hat{P}_i) \, \frac{1}{P^2_i} \, \Amp(\hat{P}_i,i+1,\cdots,2,\hat{1}) 
\nn \\
&=&
\sum_{i=1,n-3} (-1)^{n-i} \Amp(\hat{1},2,\cdots,\hat{P}_i) \, \frac{1}{P^2_i} \, (-1)^{i+2} \, \Amp(-\hat{P}_i,i,\dots,n-1,\hat{n}) 
\nn \\
&=&
(-1)^n \Amp(1,2,\dots,n-1,n) \, ,
\label{Feng_proof}
\eea
where the inductive hypothesis is used between the first and second line and which completes the proof.

For gluon amplitudes with up to 2 off-shell legs, one can always choose to shift the off-shell 
gluons in order to avoid $B$-poles, so that the only difference w.r.t. the on-shell case is the presence
of additional $n$-point amplitudes with one less off-shell leg, because of the $C$ and $D$ residues;
the latter amplitudes can be BCFW recursed themselves until all off-shell legs disappear and the color-order
reversed relation holds because of (\ref{Feng_proof}). For the $A$-terms, instead, the procedure
in (\ref{Feng_proof}) works fine, thus completing the sought generalization.

\section{Conventions}\label{AppSpinors}%

\subsection{Momenta}

We repeat here the parameterisation of the momenta used 
in~\cite{vanHameren:2014iua,vanHameren:2015bba}, which reads
\begin{align}
k_1^\mu + k_2^\mu + \cdots + k_n^\mu = 0
&\qquad\textrm{momentum conservation}\label{Eq:momcons}\\
p_1^2 = p_2^2 = \cdots = p_n^2 = 0
&\qquad\textrm{light-likeness}\label{momcon1}\\
\lid{p_1}{k_1} = \lid{p_2}{k_2}=\cdots=\lid{p_n}{k_n}=0
&\qquad\textrm{eikonal condition}
\label{Eq:eikcon}
\end{align}
where for every $k^\mu_i$ there is a corresponding orthogonal, light-like direction $p^\mu_i$.

With the help of an auxiliary light-like four-vector $q^\mu$, the momentum $k^\mu$ 
can be decomposed in terms of its light-like direction $p^\mu$, satisfying $\lid{p}{k}=0$, and a transversal part, following
\begin{equation}
k^\mu = x(q)p^\mu 
- \frac{\kapp}{2}\,\frac{\AS{p|\gamma^\mu|q}}{\SSS{pq}} 
- \frac{\kstr}{2}\,\frac{\AS{q|\gamma^\mu|p}}{\AA{qp}} \, ,
\end{equation}
with
\begin{equation}
x(q)=\frac{\lid{q}{k}}{\lid{q}{p}}
\quad,\quad
\kapp = \frac{\AS{q|\slashK|p}}{\AA{qp}}
\quad,\quad
\kstr = \frac{\AS{p|\slashK|q}}{\SSS{pq}} \, .
\end{equation}
If the momentum $k$ is on-shell, then it coincides with its associated direction, $p=k$.
The coefficients $\kapp$ and $\kstr$ do not depend on 
the auxiliary momentum $q^\mu$ (see appendix \ref{AppSchouten}) and
\begin{equation}
k^2 = -\kapp\kstr \ .
\end{equation}

We focus on the fundamental {\em color-ordered\/} or {\em dual\/} amplitudes,
where the gauge-group factors have been stripped off and all particles are massless.
By construction, these amplitudes contain only planar Feynman graphs 
and are constructed with color-stripped Feynman rules.

The polarisation vectors for gluons can be expressed as
\begin{equation}
\e^\mu_{+ } = \frac{\AS{q | \g^\mu | g}}{\sqrt{2}\AA{q g}} \, , 
\quad
\e^\mu_{-} = \frac{\AS{g |\g^\mu | q}}{\sqrt{2}\SSS{g q}} \, ,
\label{pol_gluons}
\end{equation}
where $q$ is the auxiliary light-like vector and $g$ stands for the gluon momentum.
In the expressions of the amplitudes, gluons will be denoted by the number of the corresponding particle, 
whereas we will always explicitly distinguish quarks and antiquarks with $q$ and $\qb$ respectively.

Finally, the polarization vectors for the auxiliary photons coming in pairs with the off-shell quarks are
\beq
\e^\mu_{f\,+} = \frac{\AS{q | \g^\mu | f}}{\sqrt{2}\AA{q f}} \, , 
\quad
\e^\mu_{f\,-} = \frac{\AS{f |\g^\mu | q}}{\sqrt{2}\SSS{f q}} \, ,
\label{pol_photons}
\eeq
where $f$ denotes the fermion momentum spinor and $q$ is again the auxiliary vector.

When shifting momenta with an auxiliary complex variables, as required by the BCFW procedure, 
we pick up two particles, say $i$ and $j$, and we require their light-like directions to
be the reference vector of each other, so that their momenta - with a generally 
non vanishing transverse component - become
\begin{eqnarray}
k_i^\mu = x_i(p_j)\, p_i^\mu - \frac{\kappa_i}{2}\, \frac{\AL{i} \g^\mu \SR{j}}{\SSS{ij}} - \frac{\kstr_i}{2}\, \frac{\AL{j} \g^\mu \SR{i}}{\AA{ji}}
\nn \\
k_j^\mu = x_j(p_i)\, p_j^\mu - \frac{\kappa_j}{2}\, \frac{\AL{j} \g^\mu \SR{i}}{\SSS{ji}} - \frac{\kstr_j}{2}\, \frac{\AL{i} \g^\mu \SR{j}}{\AA{ij}} \, .
\end{eqnarray}
In our notation $\SR{p_a}(\AR{p_a}) \equiv \SR{a} (\AR{a})$, where $g_a$ is any gluon, 
whereas for quarks $\SR{p_q} (\AR{p_q}) \equiv \SR{q} (\AR{q})$ and $\SR{p_{\qb}} (\AR{p_{\qb}}) \equiv \SR{\qb} (\AR{\qb})$. \\
Just as in \cite{vanHameren:2014iua}, we choose the shift vector to be
\begin{equation}
e^\mu = \frac{1}{2}\, \AL{i} \g^\mu \SR{j}  \, ,
\end{equation}
which satisfies
\beq
p_i \cdot e = p_j \cdot e = e \cdot e = 0 \, .
\eeq
Shifted quantities are denoted by a hat symbol. 
The most general case shifted momenta are thus
\bea
\hat{K}_i^\mu = k_i + z e^\mu 
&=&
x_i(p_j)\, p_i^\mu - \frac{\kappa_i - z\SSS{ij}}{2}\, \frac{\AL{i} \g^\mu \SR{j}}{\SSS{ij}} - \frac{\kstr_i}{2}\, \frac{\AL{j} \g^\mu \SR{i}}{\AA{ji}}
\nn \\
&=& 
x_i(p_j)\, p_i^\mu - \frac{\hat{\kappa}_i}{2}\, \frac{\AL{i} \g^\mu \SR{j}}{\SSS{ij}} - \frac{\kstr_i}{2}\, \frac{\AL{j} \g^\mu \SR{i}}{\AA{ji}} \, ,
\nn \\
\hat{K}_j^\mu = k_j - z e^\mu 
&=&
x_j(p_i)\, p_j^\mu - \frac{\kappa_j}{2}\, \frac{\AL{j} \g^\mu \SR{i}}{\SSS{ji}} - \frac{\kstr_j + z \AA{ij}}{2}\, \frac{\AL{i} \g^\mu \SR{j}}{\AA{ij}}
\nn \\
&=&
x_j(p_i)\, p_j^\mu - \frac{\kappa_j}{2}\, \frac{\AL{j} \g^\mu \SR{i}}{\SSS{ji}} - \frac{\kstrhat_j}{2}\, \frac{\AL{i} \g^\mu \SR{j}}{\AA{ij}} \, .
\label{shifts}
\eea
Momentum conservation and the eikonal conditions $p_i \cdot \hat{K}_i = 0$ and $p_j \cdot \hat{K}_j = 0$ are manifestly preserved.
So, for an off-shell momentum, only one of the scalar coefficients ($\kappa$ or $\kstr$) 
in the transverse momentum shifts, whereas the directions do not shift. 

We report the shifts induced by $\AL{i}\g^\mu\SR{j}/2$,
\begin{eqnarray}
e^\mu &=& \frac{1}{2}\AL{i}\g^\mu\SR{j} 
\Leftrightarrow 
\left\{ \begin{array}{c}
\textrm{$i$ off-shell:} \quad \kapphat_i = \kappa_i - z \SSS{ij} 
\\
\textrm{$i$ on-shell:} \quad \SR{\hat{i}} = \SR{i} + z \SR{j} 
\\
\textrm{$j$ off-shell:} \quad \kstrhat_i = \kstr_j + z \AA{ij} 
\\
\textrm{$j$ on-shell:} \quad \AR{\hat{j}} = \AR{j} - z \AR{i}
\end{array} \right.
\label{shift1}
%
\end{eqnarray}
%
%
%
%
%

\subsection{Spinors}

The left- and right-handed spinors used in this paper can be defined in the following way:
\begin{eqnarray}
\SR{p}
&=& 
\begin{pmatrix} L(p) \\ \mathbf{0} \end{pmatrix}
\quad \quad
L(p) = \frac{1}{\sqrt{|p^0+p^3|} }\, \begin{pmatrix} -p^1+ \imag\, p^2 \\ p^0+p^3 \end{pmatrix} 
\nn \\
\AR{p}
&=& 
\begin{pmatrix} \mathbf{0} \\ R(p) \end{pmatrix}
\quad \quad
R(p) = \frac{\sqrt{|p^0 + p^3|} }{ p^0 + p^3 } \, \begin{pmatrix} p^0 + p^3 \\ p^1+ \imag \, p^2 \end{pmatrix}  
\end{eqnarray}
The "dual" spinors are defined as
\begin{equation}
\SL{p} = \big(\, (\mathcal{E}L(p))^T \, ,\,\mathbf{0} \,\big)
\quad \quad
\AL{p} = \big(\, \mathbf{0} \, \, ( \mathcal{E}^T R(p) )^T \,\big)
\quad \quad  \textrm{where} \quad
\mathcal{E} = \begin{pmatrix} 0 & 1 \\ -1 & 0 \end{pmatrix}
\end{equation}
The definition of the ``dual'' spinors does not involve complex conjugation and all spinors are well defined for complex momenta. \\
Defining the Pauli vector as
\begin{equation}
\sigma^\mu = \left( \mathbf{1}_{2\times 2}, \stackrel{\rightarrow}{\sigma} \right) \equiv \left( \sigma^0, \sigma^1,\sigma^2,\sigma^3 \right) \, ,
\end{equation}
our conventions for the Dirac matrices are
\begin{equation}
\gamma^0 = 
\left(
\begin{array}{cc}
            0 & \sigma^0 \\
\sigma^0 & 0
\end{array}
\right)
\, , \quad
\gamma^i = 
\left(
\begin{array}{cc}
            0 & \sigma^i \\
- \sigma^i & 0
\end{array}
\right) \, , 
\quad
i = 1,2,3 \, .
\end{equation}
so that $\gamma^5$ is given by
\begin{equation}
\gamma^5 \equiv \imag\, \gamma^0 \gamma^1\gamma^2\gamma^3 = 
\left(
\begin{array}{cc}
-\mathbf{1} & 0 \\
0 & \mathbf{1}
\end{array}
\right)
\, .
\end{equation}
The dyadic product satisfies the identity
\begin{equation}
\AR{p}\SL{p} + \SR{p}\AL{p} = \slashp = \gamma_\mu p^\mu
\end{equation}
The following spinor identities hold and are extensively used in the paper, with the convention that $p^\mu$, $q^\mu$ and $r^\mu$ are always light like,
whereas $k^\mu$ does not have to be light-like
\begin{eqnarray}
\AL{p}\SR{q} &=& \SL{p}\AR{q} = 0 \nn
\\
\AL{p}\AR{p} &=& \SL{p}\SR{p} = 0 \nn
\\
\slashp\AR{p} &=& \slashp\SR{p} = \AL{p}\slashp = \SL{p}\slashp = 0 \nn
\\
\AL{p}\AR{q} &\equiv& \AA{pq} = - \AA{qp} \nn
\\
\SL{p}\SR{q} &\equiv& \SSS{pq} = -\SSS{qp}  \nn
\\
\AA{pq}\SSS{qp} &=& 2\lid{p}{q} \nn
\\
\AS{p|\slashr|q} &=& \AA{pr}\SSS{rq} \nn
\\
\gamma_\mu \AL{p} \gamma^\mu \SR{q}  &=& 2\, \left( \SR{q}\AL{p} + \AR{p}\SL{q}   \right)
\nn
\\
\AL{r}\gamma_\mu \SR{s} \AL{p} \gamma^\mu \SR{q}  &=& 2\, \AA{r p} \SSS{q s}
\nn
\\
\AL{p}\gamma^{\mu_1}\dots \gamma^{\mu_n}\SR{q} &=& \SL{q}\gamma^{\mu_n}\dots \gamma^{\mu_1}\AR{p}
\nn\\
\AL{p} \gamma^{\mu_1}\dots \gamma^{\mu_n} \AR{q} &=& -  \AL{q} \gamma^{\mu_n} \dots \gamma^{\mu_1} \AR{p}
\nn\\
\SL{p} \gamma^{\mu_1}\dots \gamma^{\mu_n} \SR{q} &=& -  \SL{q} \gamma^{\mu_n} \dots \gamma^{\mu_1} \SR{p}
\end{eqnarray}
The last two relations trivially reduce to $0=0$ for odd $n$, and the one before reduces to $0=0$ for even $n$.

\subsection{Schouten identity}\label{AppSchouten}

A very insightful way to express the Schouten identity is to write it as a completeness relation.
For any $p^\mu,r^\mu$ with $p^2=r^2=0$ and $\lid{p}{r}\neq0$ we have
\begin{equation}
\frac{\AR{r}\AL{p}}{\AA{pr}} + \frac{\AR{p}\AL{r}}{\AA{rp}} + \frac{\SR{r}\SL{p}}{\SSS{pr}} + \frac{\SR{p}\SL{r}}{\SSS{rp}} = 1 \, .
\label{Schouten}
\end{equation}
A simple application is the proof that $\kappa$ and $\kstr$ are independent of the auxiliary momentum.
Inserting the identity operator strategically, we find for $\kappa$ 
\begin{equation}
\kappa = \frac{\AS{q|\slashK|p}}{\AA{qp}} 
= \frac{\AS{q|1\slashK|p}}{\AA{qp}}
= \frac{\AA{qr}\AS{p|\slashK|p}}{\AA{pr}\AA{qp}} + \frac{\AA{qp}\AS{r|\slashK|p}}{\AA{rp}\AA{qp}}
= \frac{\AS{r|\slashK|p}}{\AA{rp}} \, ,
\end{equation}
where we used the fact that $\AS{p|\slashK|p}=2\lid{p}{k}=0$.
The same can be shown for $\kstr$.

\section{Some explicit calculations}\label{Calculus}

This Appendix is devoted to preset some explicit BCFW derivations 
which were omitted in the text for the sake of shortness.

\subsection{$\Amp(1^+,\qb^{*+},q^{*-})$}

\begin{figure}[ht]
\begin{center}
\includegraphics[scale=0.80]{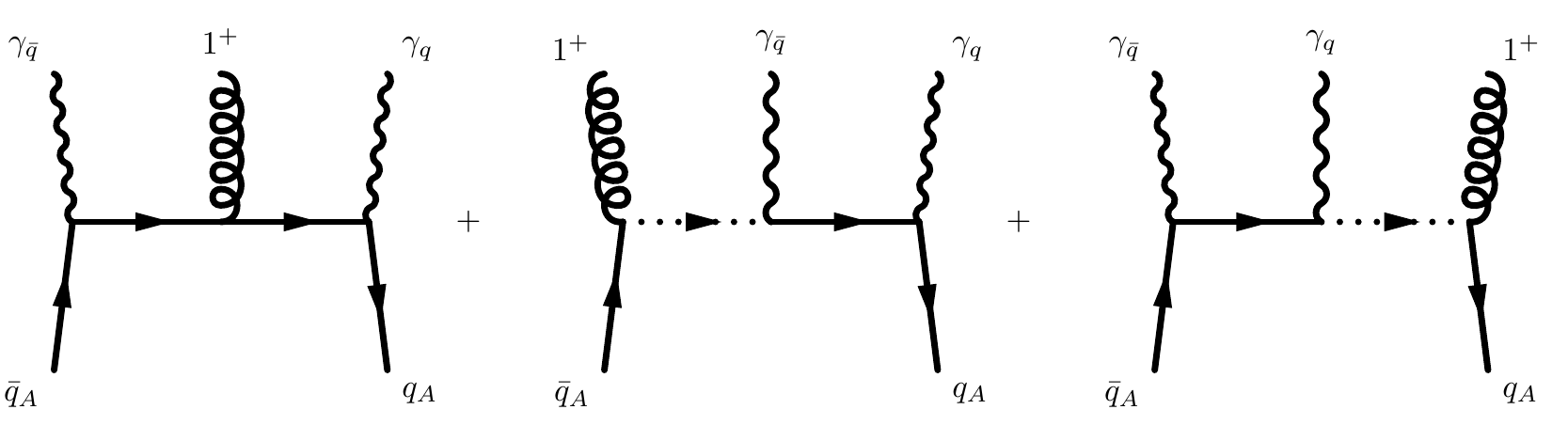}
\caption{Feynman diagrams contributing to $\Amp(1^+,\qb^{*+},q^{*-})$}
\label{off_qqb}
\end{center}
\end{figure}

The first calculation we present is the evaluation of the first of the three-point amplitudes in (\ref{3ferfer}).
We do this in two ways, diagrammatically (see Fig~\ref{off_qqb}) and via BCFW, in order to provide an
explicit cross-check. 
There are three Feynman diagrams contributing, shown in Fig. \ref{off_qqb}, where all momenta are taken as incoming.
In the first one the two off-shell fermions are coming from the QED
vertex between the auxiliary fermion and photon, in the remaining two
auxiliary particles pair are bent open in turn and
auxiliary fermions appear in internal lines as eikonal propagators,
denoted with a dashed line according to the Feynman rules~\cite{vanHameren:2015bba}.
Notice that, as pointed out in \cite{vanHameren:2013csa}, auxiliary photons interact only with the
corresponding auxiliary fermion, not with any other charged fermion. 

The implicit expression for our amplitude that one gets from the Feynman diagrams is thus
\beq
\Amp(1^+,\qb^{*+},q^{*-}) = \AL{q}   
\frac{\slasheps_{q\, +}}{\sqrt{2}} \frac{-\slashk_q}{k_q^2}  \frac{\slasheps_{1\,+}}{\sqrt{2}} \frac{\slashk_{\qb}}{k_{\qb}^2}  \frac{\slasheps_{\qb\,-}}{\sqrt{2}}  +
\frac{\slasheps_{q\,+}}{\sqrt{2}} \frac{\slashk_q}{k_q^2}  \frac{\slasheps_{\qb\,-}}{\sqrt{2}} \frac{\slashp_{\qb}}{p_{\qb}\cdot k_q}  \frac{\slasheps_{1\,+}}{\sqrt{2}} +
\frac{\slasheps_{1\,+}}{\sqrt{2}} \frac{\slashp_{q}}{p_q\cdot k_{\qb}}  \frac{\slasheps_{q\,+}}{\sqrt{2}} \frac{\slashk_{\qb}}{k_{\qb}^2}  \frac{\slasheps_{\qb\,-}}{\sqrt{2}} 
\SR{\qb} \, ,
\eeq
where the helicity of the auxiliary photons is enforced by their position, because (see Eq. (\ref{pol_photons}) )
the other one would give just zero. We choose the fermion momenta $p_q$ and $p_{\qb}$ to be auxiliary vectors
for each other's auxiliary photon.
One degree of freedom still at our disposal in the computation is the choice of the auxiliary vector for
the polarization of the gluon. From Eq. (\ref{pol_gluons}) one can check that choosing $p_q$ 
annihilates the first and third contribution, so that we are left with
\bea
&&
\frac{\AA{q\qb} \, \AL{\qb} \, \slashk_{q} \SR{q} \SSS{q\qb} \, \AA{\qb q} \, \SSS{q \qb} \, \SSS{1\qb} }{\AA{\qb q} \, (- \kapp_q \kstr_{q}) \, \AL{\qb} \slashk_{q} \SR{\qb} \, \SSS{\qb q} \, \AA{q1}}
=
\frac{\AA{q\qb}^2\, \SSS{1\qb} \, \kapp_q \AA{\qb q} }{ \kapp_q \kstr_{\qb} \, \AA{1\qb}\, \SSS{1\qb}\, \AA{1q} } = 
\frac{1}{\kapp_q}\, \frac{\AA{q\qb}^2}{\AA{1q}\, \AA{1\qb}}
\, ,
\eea
after exploiting the definitions of $\kapp_q$ and $\kstr_q$ and momentum conservation.

\begin{figure}[ht]
\begin{center}
\includegraphics[scale=0.80]{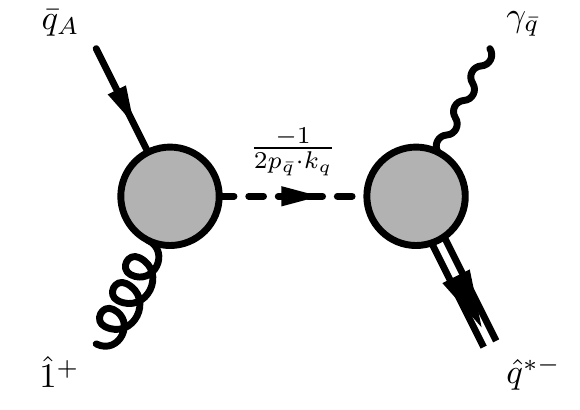}
\caption{The only BCFW residue for $\Amp(1^+,\qb^{*+},q^{*-})$ with shift vector $e^\mu = \AL{q}\g^\mu\SR{1}/2$}
\label{3pt_1qqb}
\end{center}
\end{figure}

Now we compute the same amplitude by BCFW recursion using $\AL{q}\g^\mu\SR{1}/2$ as shift vector, which respects the LW
prescription in the on-shell limit, so that we only have a $B$-pole, due to the eikonal propagator of the auxiliary eikonal fermion
used for $\qb$ going on-shell, as depicted in Fig. \ref{3pt_1qqb}.
The following relations are derived by requiring the location of the pole to be $p_{\qb} \cdot \hat{K}_q = 0$,
\beq
z = \frac{\AA{1\qb}}{q\qb} \, ,
\quad
\hat{\kapp_q} = \frac{\AL{\qb} \slashp_1 + \slashk_q \SR{q} }{ \AA{\qb q}} = \kstr_{\qb} \frac{\SSS{\qb q}}{\AA{q\qb}} \, ,
\quad
\AR{\hat{1}} = \frac{\AA{1q}}{\AA{\qb q}}\, \AR{\qb} \, .
\label{shifts3offqqb}
\eeq
In particular, they imply $\hat{\epsilon}_{1\,+}^\mu = \epsilon_{1\,+}^\mu$, giving
\beq
\Amp(1^+,\qb^{*+},q^{*-}) = \AL{\qb} \frac{\slasheps_{1\,+}}{\sqrt{2}} \SR{\qb} \, \frac{1}{\AA{1\qb}\SSS{\qb1}} \, 
\AL{q}  \frac{\slasheps_{q\,+}}{\sqrt{2}}\, \frac{\hat{K}_q}{\hat{K}_q^2} \, \frac{\slasheps_{\qb\,-}}{\sqrt{2}}  \SR{\qb}   \, .
\label{3offqqb_prel}
\eeq
On the left of the unshifted eikonal propagator is the 2-point amplitude of the gluon and the $\qb$ fermion current, 
whereas on its right is the product of the two auxiliary photon vertices with the $q$ propagator 
( evaluated for shifted momentum ) in the middle. One can visualise these "unconventional" 2-point 
amplitudes thinking about them as the parts of the second Fenman diagram in Fig. \ref{3pt_1qqb} 
on the left and right side of the propagator and evaluated for shifted momenta.

Replacing the relations (\ref{shifts3offqqb}) into (\ref{3offqqb_prel}) one finally gets
\bea
&&
\frac{ \AA{\qb q} \, \SSS{1\qb} \AL{\qb}  \hat{\slashK}_q \SR{q} }{\AA{1\qb}\,\SSS{\qb 1}\,\AA{1q}\, \hat{\kapp_q}\, \kstr_q } 
= 
\frac{\AA{\qb q} \, \SSS{1\qb} \, \AL{\qb} - \slashk_{\qb} - \hat{\slashp}_1 \SR{q} }{ {\AA{1\qb}\,\SSS{\qb 1}\,\AA{1q}\, \hat{\kapp_q}\, \kstr_q}} =
\frac{ - \AA{\qb q} \, \SSS{1\qb} \, \kstr_{\qb} \, \SSS{\qb q} }{\AA{1q} \, \AA{1\qb} \, \SSS{\qb q}\, \kstr_{\qb} \, \SSS{\qb q} / \AA{q\qb}}=
\frac{1}{\kstr_q} \, \frac{\AA{q\qb}^2}{\AA{1q} \, \AA{1\qb}}  \, ,
\nn \\
\eea
as announced.
For all the other amplitudes exactly the same procedure can be repeated and they can be derived
in two independent ways, thus providing a non trivial cross-check.

\subsection{$\Amp(2^*,\qb^+,q^-,1^*) $}%

Here we explicitly compute the first amplitude in \ref{gstargstar}.
The other one is obtained similarly.
We choose to shift the gluons through $e^\mu = \AL{1}\gamma^\mu\SR{2}/2$, so that we have
3 contributions: one $C$, one $D$ and one ordinary $A$ poles,
\bea
\Amp(2^*,\qb^+,q^-,1^*) 
&=& 
\frac{1}{x_1\kapp_1}\, \Amp(\hat{2}^*,\qb^+,q^-,\hat{1}^+) +
\frac{1}{x_2\kstr_2}\, \Amp(\hat{2}^-,\qb^+,q^-,\hat{1}^*) 
\nn \\
&+& 
\Amp(\hat{2}^*,\qb^+,\hat{K}^-)\,\frac{1}{(k_2 + p_{\qb})^2}\, \Amp(-\hat{K}^+,q^-,\hat{1}^*) 
\label{First_prelim}
\eea
The first two terms can be worked out by looking at the expressions for the MHV amplitudes with one off-shell gluon and one
fermion pair computed in \cite{vanHameren:2015bba}, i.e.
\bea
\Amp(1^*,\qb^+,q^-,2^+,\dots,n^+) 
&=& 
\frac{1}{\kstr_1}\, \frac{ \AA{1 q}^3 \AA{1\qb} }{\AA{1 \qb} \AA{\qb q} \dots \AA{n-1 |n} \AA{ n 1} } \, ,
\nn \\
\Amp(1^*,\qb^+,q^-,2^-,\dots,n^-) 
&=& 
\frac{1}{\kapp_1}\, \frac{ \SSS{1\qb}^3 \AA{1q} }{\SSS{1n} \SSS{n |n-1} \dots \SSS{q\qb}\SSS{\qb1} } \, ,
\eea
replacing them into (\ref{First_prelim}) and using the shifted quantities of Table \ref{CDg},
obtaining after a little algebra
\bea
\frac{1}{x_2\kstr_2}\, \Amp(\hat{2}^-,\qb^+,q^-,\hat{1}^*) 
&=&
\frac{1}{\kstr_2} \, \frac{\AA{21}^3 \, \SSS{1\qb}^3}{ \AL{2} \slashk_2 + \slashk_1 \SR{1} \, \AL{1} \slashk_2 \SR{\qb} \, \SL{1} \slashk_2 \SR{1} \SSS{\qb q} } \, ,
\nn \\
\frac{1}{x_1\kapp_1}\, \Amp(\hat{2}^*,\qb^+,q^-,\hat{1}^+)
&=&
\frac{1}{\kapp_1} \, \frac{\AA{2q}^3 \, \SSS{12}^3 }{  \AL{2} \slashk_2 + \slashk_1 \SR{1} \, \AL{q} \slashk_1 \SR{2} \, \SL{2} \slashk_1 \SR{2} \AA{\qb q} } \, .
\eea
As for the ordinary $A$ pole, its location is given by
\beq
(k_2 + p_{\qb} + z\,e )^2 = 0 \Rightarrow z = - \frac{(k_2+p_{\qb})^2}{2\, e\cdot \left( k_2+p_{\qb} \right)} 
\Rightarrow 
\hat{\slashK} = \slashk_2 + \slashp_{\qb} + \frac{(k_2+p_{\qb})^2\, \slashe }{\AL{1}  \slashp_q + \slashk_1 \SR{2}}
\eeq
which gives
\bea
&&
\Amp(\hat{2}^*,\qb^+,\hat{K}^-)\,\frac{1}{(k_2+ p_{\qb})^2}\, \Amp(-\hat{K}^+,q^-,\hat{1}^*)  =
\nn \\
&& 
\frac{1}{\kapp_2}\, \frac{ \SSS{2\qb}^3\SSS{2\hat{K}}}{\SSS{\qb2} \, \SSS{2\hat{K}} \, \SSS{\hat{K} \qb}} \, 
\frac{1}{(k_2+p_{\qb})^2}  \,
\frac{1}{\kstr_1} \, \frac{\AA{1q}^3\AA{1\hat{K}}}{\AA{\hat{K}q} \, \AA{q1} \, \AA{1\hat{K}}} =
- \frac{1}{\kstr_2\,\kapp_1} \frac{1}{(k_2+p_{\qb})^2}\, \frac{\SSS{2\qb} \AA{1q}^2}{ \AL{q} \hat{\slashK} \SR{\qb}}
\nn \\
&=&
- \frac{1}{\kstr_1\,\kapp_2} \frac{1}{(k_2+p_{\qb})^2} \, 
\frac{ \SSS{2\qb}^2 \, \AA{1q}^2 \, \AL{1} \slashk_2+\slashp_{\qb} \SR{2} }{\AL{q} \slashk_2 \SR{\qb} \, \AL{1} \slashk_2 + \slashp_{\qb} \SR{2} + (k_2+p_{\qb})^2\, \SSS{\qb2} \, \AA{q1}} \, .
\eea
Notice that in the $A$ residue we have used, among the 2 possible representation of the 3-point amplitudes,
the one which would have been enforced in the on-shell case by the helicities demanding the shift vector $e^\mu=\AL{1}\gamma^\mu\SR{2}/2$.
So we can write down the sought amplitude
\bea 
\Amp(2^*,\qb^+,q^-,1^*) 
&=& 
\frac{1}{\kapp_1} \, \frac{\AA{2q}^3 \, \SSS{12}^3 }{ \AA{\qb q} \, \AL{2} \slashk_2 + \slashk_1 \SR{1} \, \AL{q} \slashk_1 \SR{2} \, \AL{2} \slashk_1 \SR{2} }
\nn \\
&+&
\frac{1}{\kstr_2} \, \frac{\AA{21}^3 \, \SSS{1\qb}^3}{ \SSS{\qb q} \, \AL{2} \slashk_2 + \slashk_1 \SR{1} \, \AL{1} \slashk_2 \SR{\qb} \, \AL{1} \slashk_2 \SR{1} }
\nn \\
&-&
\frac{1}{\kstr_1\,\kapp_2} \, \frac{1}{(k_2+p_{\qb})^2} \, 
\frac{ \SSS{2\qb}^2 \, \AA{1q}^2 \, \AL{1} \slashk_2+\slashp_{\qb} \SR{2} }{\AL{q} \slashk_2 \SR{\qb} \, \AL{1} \slashk_2 + \slashp_{\qb} \SR{2} + (k_2+p_{\qb})^2\, \SSS{\qb2} \, \AA{q1}} \, .
\eea
%

\subsection{$\Amp(2^*,\qb^+,q^*,1^-)$}%

In order to compute $\Amp(2^*,\qb^+,q^*,1^-)$ we choose the vector $e^\mu = \AL{q}\g^\mu\SR{2}/2$, so that
the $C$-residue vanishes. This implies that we have in principle three contributions, as we had for $\Amp(2^*,\qb^+,q^*,1^+) $, 
\bea
\Amp(2^*,\qb^+,q^*,1^-) 
&=&
\frac{1}{x_2\,\kstr_2} \, \Amp(\hat{2}^-,\qb^+,q^*,1^-) 
\nn \\
&+& 
\Amp(2^*,\qb^+,\hat{K}^-) \, \frac{1}{(k_2+p_{\qb})^2} \, \Amp(-\hat{K}^+,q^*,1^-) 
\nn \\
&+& 
\Amp(1^-, 2^*,\hat{K}^+) \, \frac{1}{( p_1 + k_2)^2} \, \Amp(-\hat{K}^-,\qb^+,q^*)  \, .
\eea
Let us recall the general structure of the so-called subleading amplitudes ($n$ gluons with the same helicity 
plus one fermion pair with one off-shell member) from \cite{vanHameren:2015bba}:
\bea
\Amp(1^+,2^+,\dots,n^+,\bar{q}^*,q^-) 
&=& 
\frac{ \AA{\qb q}^3}{\AA{12}\AA{23}\dots\AA{n\qb}\AA{\qb q}\AA{q 1}}  \, ,
\nn \\
\Amp(1^-,2^-,\dots,n^-,\bar{q}^*,q^-) 
&=&
\frac{ \SSS{q\qb}^3}{\SSS{1q}\SSS{q\qb}\SSS{\qb n}\dots\SSS{32}\SSS{21}}  \, .
\label{subleading}
\eea
After inserting (\ref{subleading}) and the expressions for the 3-point amplitudes 
given in appendix \ref{3less}, we find after some algebra
\bea
\Amp(2^*,\qb^+,q^*,1^-) 
&=&
\frac{1}{x_2\,\kstr_2} \, \frac{\SSS{q\qb}^3}{\SSS{1q}\SSS{q\qb}\SSS{\qb \hat{2}} \SSS{\hat{2}1}} 
\nn \\ 
&& \hspace{-12mm }
- \frac{1}{\kapp_2\,\kstr_q}\, \left(
\frac{1}{(k_2+p_{\qb})^2} \, \frac{ \SSS{2\qb}^2 \AA{1q}^2 }{ \AL{q} \hat{\slashK} \SR{\qb} } - 
\frac{1}{(p_1 + k_2)^2} \,\frac{ \AL{q} \hat{\slashK} \SR{2}^3 }{\SSS{12} \AA{q\qb} \AL{q} \hat{\slashK} \SR{1} } \,
\right) \, .
\eea
A look at Table \ref{CDg} allows to compute the first term, whereas for the remaining two it is not necessary
to compute the exact location of the pole, as the spinors between which $\hat{K}$ is sandwiched annihilate 
$\slashe = \AR{q}\SL{2} + \SR{2}\AL{q} $. This leads to the final result, together with its parity conjugated
\bea
\Amp(2^*,\qb^+,q^*,1^-) 
&=&
\frac{1}{\kstr_2} \, \frac{\SSS{q\qb}^2 \, \AA{2q}^2 }{ \SSS{q1} \, \AL{q} \slashk_2  \SR{1} \, \AL{q} \slashk_2 \SR{\qb} } 
\nn \\ 
&& \hspace{-12mm }
- \frac{1}{\kapp_2\,\kstr_q}\, \left(
\frac{1}{(k_2 + p_{\qb})^2} \, \frac{ \SSS{2\qb}^2 \, \AA{1q}^2 }{ \AL{q} \slashk_2 \SR{\qb} } - 
\frac{1}{(p_1 + k_2)^2} \,\frac{ \AL{q} \slashk_2 + \slashp_1 \SR{2}^3 }{ \SSS{12} \, \AA{q\qb}  \, \AL{q} \slashk_2 \SR{1} }
\right) \, ,
\nn \\
\Amp(2^*,\qb^-,q^*,1^+) 
&=&
\frac{1}{\kapp_2} \, \frac{\AA{q\qb}^2 \, \SSS{2q}^2 }{ \AA{1q} \, \AL{1} \slashk_2  \SR{q} \, \AL{\qb} \slashk_2 \SR{q} } 
\nn \\ 
&& \hspace{-12mm }
- \frac{1}{\kstr_2\,\kapp_q}\, \left(
\frac{1}{(k_2 + p_{\qb})^2} \, \frac{ \AA{2\qb}^2 \, \SSS{1q}^2 }{ \AL{\qb} \slashk_2 \SR{q} } - 
\frac{1}{(p_1 + k_2)^2} \,\frac{ \AL{2} \slashk_2 + \slashp_1 \SR{q}^3 }{ \AA{12} \, \SSS{q\qb}  \, \AL{1} \slashk_2 \SR{q} }
\right) \, .
\eea
%

\subsection{$\Amp(2^+,\qb^{*+},q^{*-},1^-)$}

For $\Amp(2^+,\qb^{*+},q^{*-},1^-)$ we obviously choose $e^\mu = \AL{1}\gamma^\mu\SR{2}/2$
and we get one $A$ and two $B$ residues,
\bea
\Amp(2^+,\qb^{*+},q^{*-},1^-) 
&=&
\Amp(2^+,\qb^{*+},\hat{K}^-) \, \frac{1}{(p_2+k_{\qb})^2} \, \Amp(-\hat{K}^+,q^{*-},1^-)
\nn \\
&-&
\Amp_2(\hat{2}^+,\hat{\qb}^{*-}) \, \frac{1}{2\,p_{\qb}\cdot p_2} \, \Amp(\hat{\qb}^{*+},q^{*-},\hat{1}^-) 
\nn \\
&-& 
\Amp(\hat{2}^+,\qb^{*+},\hat{q}^{*-}) \, \frac{1}{2\,p_q\cdot p_1} \, \Amp_2(\hat{q}^{*-},\hat{1}^-) 
\nn \\
&=&
\frac{1}{\kapp_{\qb}} \, \frac{ \SSS{2\qb}^3 \, \SSS{2\hat{K}} }{\SSS{2\hat{K}} \, \SSS{\hat{K}\qb} \, \SSS{\qb2} } \, \frac{1}{(p_2+k_{\qb})^2} \, 
\frac{1}{\kstr_q} \, \frac{\AA{1q}^3 \, \AA{1\hat{K}}}{ \AA{1\hat{K}} \, \AA{\hat{K}q} \, \AA{q1} } 
\nn \\
&+&
\frac{\AA{\qb1}}{\AA{\qb2}\AA{21}} \, \frac{1}{\hat{\kapp}_{\qb}} \, \frac{\SSS{q\qb}^3}{\SSS{\hat{1}q} \, \SSS{q\qb} \, \SSS{\qb\hat{1}}} + 
\frac{\SSS{2q}}{\SSS{q1} \, \SSS{12}} \, \frac{1}{\hat{\kstr}_q} \, \frac{\AA{q\qb}^3}{\AA{\hat{2}\qb} \, \AA{\qb q} \, \AA{q\hat{2}}} \, .
\eea
For the first term we find the following pole location and shifted momentum,
\beq
z = - \frac{(p_2+k_{\qb})^2}{\AL{1} \slashk_{\qb} \SR{2}} \Rightarrow \hat{\slashK} = 
\slashp_2+\slashk_{\qb} - \frac{(p_2+k_{\qb})^2}{\AL{1} \slashk_{\qb} \SR{2}} \, \left(  \AR{1}\SL{2} + \SR{2}\AL{1} \right) \, ,
\eeq
from which we get
\bea
&&
\frac{1}{\kapp_{\qb}} \, \frac{ \SSS{2\qb}^3 \, \SSS{2\hat{K}} }{\SSS{2\hat{K}} \, \SSS{\hat{K}\qb} \, \SSS{\qb2} } \, \frac{1}{(p_2+k_{\qb})^2} \, 
\frac{1}{\kstr_q} \, \frac{\AA{1q}^3 \, \AA{1\hat{K}}}{ \AA{1\hat{K}} \, \AA{\hat{K}q} \, \AA{q1} } 
\nn \\
&=&
\frac{1}{\kappa_{\qb}\,\kstr_{q}} \, 
\frac{1}{(p_2+k_{\qb})^2} \, 
\frac{\SSS{2\qb}^2 \, \AA{1q}^2 \, \AL{1} \slashk_{\qb} \SR{2} }{\AL{q} \slashp_2 + \slashk_{\qb} \SR{\qb} \, \AL{1} \slashk_{\qb} \SR{2} + (p_2+k_{\qb})^2 \AA{1q}\SSS{2\qb}} \, .
\eea
For the second and third term, it takes the formulas collected in section \ref{res} to respectively come up with
\bea
\frac{\AA{\qb1}}{\AA{\qb2}\AA{21}} \, \frac{1}{\hat{\kapp}_{\qb}} \, \frac{\SSS{q\qb}^3}{\SSS{\hat{1}q} \, \SSS{q\qb} \, \SSS{\qb\hat{1}}} 
&=&
\frac{ \AA{1\qb}^4 \, \SSS{q\qb}^2 }{\AA{12} \, \AA{2\qb} \, \AL{1} \slashk_q \, \slashp_{\qb} \, (\slashp_1 + \slashp_2) \SR{q} \, \AL{\qb} \slashp_1 + \slashp_2 \SR{\qb} }
\nn \\ 
\frac{\SSS{2q}}{\SSS{q1} \, \SSS{12}} \, \frac{1}{\hat{\kstr}_q} \, \frac{\AA{q\qb}^3}{\AA{\hat{2}\qb} \, \AA{\qb q} \, \AA{q\hat{2}}}
&=&
\frac{ \SSS{2q}^4 \, \AA{q\qb}^2 }{\SSS{12} \, \SSS{1q} \, \AL{\qb} (\slashp_1 + \slashp_2) \, \slashp_q \, \slashk_{\qb}  \SR{2} \, \AL{q} \slashp_1 + \slashp_2 \SR{q} } \, ,
\eea
so that we finally obtain
\bea
\Amp(2^+,\qb^{*+},q^{*-},1^-) 
&=&
\frac{1}{\kappa_{\qb}\,\kstr_{q}} \, 
\frac{1}{(p_2+k_{\qb})^2} \, 
\frac{\SSS{2\qb}^2 \, \AA{1q}^2 \, \AL{1} \slashk_{\qb} \SR{2} }{ \AL{q} \slashp_2 + \slashk_{\qb} \SR{\qb} \, \AL{1} \slashk_{\qb} \SR{2} + (p_2+k_{\qb})^2 \, \AA{1q} \, \SSS{2\qb}}
\nn \\
&+& 
\frac{ \AA{1\qb}^4 \, \SSS{q\qb}^2 }{ \AA{12} \, \AA{2\qb} \, \AL{1} \slashk_{\qb} + \slashp_2 \SR{\qb} \, \AL{\qb} \, \slashp_1 + \slashp_2 \SR{q} \, \AL{\qb} \slashp_1 + \slashp_2 \SR{\qb} }
\nn \\
&+&
\frac{ \SSS{2q}^4 \, \AA{q\qb}^2 }{ \SSS{12} \, \SSS{1q} \, \AL{q} \slashk_q + \slashp_1 \SR{2} \, \AL{\qb} \slashp_1 + \slashp_2 \SR{q} \, \AL{q} \slashp_1 + \slashp_2 \SR{q} } \, .
\eea
%

\subsection{$\Amp(2^+,\qb^{*-},q^{*+},1^-)$}

For the other $2$ amplitudes in which, in the on-shell limit, the gluons and fermions with the same helicity sign are not adjacent, 
we have to make a similar procedure, but there will be no $A$ residues.
We show explicitly the calculation of $\Amp(2^+,\qb^{*-},q^{*+},1^-)$ with $e^\mu = \AL{1} \gamma^\mu \SR{2}/2$,
which is similar for the adjoint $\Amp(2^-,\qb^{*+},q^{*-},1^+)$ with $e^\mu = \AL{2} \gamma^\mu \SR{1}/2$.
\bea
\Amp(2^+,\qb^{*-},q^{*+},1^-) 
&=&
\Amp(2^+,\hat{\qb}^{*-},\hat{K}^+) \, \frac{1}{(p_2+k_{\qb})^2} \, \Amp(-\hat{K}^-,q^{*+},1^-)
\nn \\
&-&
\Amp_2(\hat{2}^+,\hat{\qb}^{*-}) \, \frac{1}{2\,p_{\qb}\cdot p_2} \, \Amp(\hat{\qb}^{*-},q^{*+},\hat{1}^-) 
\nn \\
&-&
\Amp(\hat{2}^+,\qb^{*-},\hat{q}^{*,+}) \, \frac{1}{2\,p_q \cdot p_1} \, \Amp_2(\hat{q}^{*+},\hat{1}^-)
\nn \\
&=&
\frac{\AA{\qb1}}{\AA{\qb2} \, \AA{21}} \, \frac{1}{\kapp_q} \, \frac{\SSS{q\qb}^3}{\SSS{\hat{1}q} \, \SSS{q\qb} \, \SSS{\qb\hat{1}}} +
\frac{\SSS{2q}}{\SSS{q1} \, \SSS{12}} \, \frac{1}{\kstr_{\qb}} \, \frac{\AA{q\qb}^3}{\AA{\hat{2}\qb} \, \AA{\qb q} \, \AA{q\hat{2}} } 
\eea
The $A$-residue is clearly zero, whereas for the other two we use again \ref{res} and get
\bea
\frac{\AA{\qb1}}{\AA{\qb2} \, \AA{21}} \, \frac{1}{\kapp_q} \, \frac{\SSS{q\qb}^3}{\SSS{\hat{1}q} \, \SSS{q\qb} \, \SSS{\qb\hat{1}}}
&=&
\frac{1}{\kapp_q} \, \frac{\AA{1\qb}^3 \, \SSS{q\qb}^2}{\AA{12} \, \AA{2\qb} \, \AL{\qb} (\slashp_1+\slashp_2) \, \slashp_{\qb} \, (\slashp_1+\slashp_2)  \SR{q}}
\nn \\
\frac{\SSS{2q}}{\SSS{q1} \, \SSS{12}} \, \frac{1}{\kstr_{\qb}} \, \frac{\AA{q\qb}^3}{\AA{\hat{2}\qb} \, \AA{\qb q} \, \AA{q\hat{2}} } 
&=&
\frac{1}{\kstr_{\qb}} \, \frac{\SSS{2q}^3 \, \AA{q\qb}^2}{\SSS{12} \, \SSS{q1} \, \AL{\qb} (\slashp_1+\slashp_2) \, \slashp_q \, (\slashp_1+\slashp_2)  \SR{q}} \, ,
\eea
so that the final result is
\bea
\Amp(2^+,\qb^{*-},q^{*+},1^-) 
&=&
\frac{1}{\kapp_q} \, \frac{\AA{1\qb}^3 \, \SSS{q\qb}^2}{\AA{12} \, \AA{2\qb} \, \AL{\qb} \, \slashp_1+\slashp_2  \SR{q} \, \AL{\qb} \slashp_1+\slashp_2 \SR{\qb} } 
\nn \\
&+&
\frac{1}{\kstr_{\qb}} \, \frac{\SSS{2q}^3 \, \AA{q\qb}^2}{\SSS{12} \, \SSS{q1} \, \AL{\qb} \slashp_1+\slashp_2 \SR{q} \, \AL{q} \slashp_1+\slashp_2 \SR{q}} \, .
\eea

\end{document}